%% file: template.tex
\documentclass[50pt]{elsarticle}
\graphicspath{ {./figures/} }
\usepackage{hyperref}
\usepackage{float}
\usepackage{verbatim} 
\usepackage{apalike}
\usepackage{graphicx}
\usepackage{color}
\usepackage{amsmath}
\usepackage{amssymb}
\usepackage{subfig}
\usepackage{url}
\usepackage{arydshln}
\usepackage{enumitem}
\usepackage{xcolor}

\restylefloat{figure}
\restylefloat{table}

\journal{Expert Systems with Applications}

\biboptions{authoryear}

\begin{document}
\begin{frontmatter}










\title{Combining piano performance dimensions \\for score difficulty classification}

\author[label1]{Pedro Ramoneda \corref{cor1}}
\ead{pedro.ramoneda@upf.edu}

\author[label2]{Dasaem Jeong}
\ead{dasaemj@sogang.ac.kr}

\author[label1]{Vsevolod Eremenko}
\ead{vsevolod.eremenko@upf.edu}

\author[label1]{Nazif Can Tamer}
\ead{nazifcan.tamer@upf.edu}

\author[label1]{Marius Miron}
\ead{marius.miron@upf.edu}

\author[label1]{Xavier Serra}
\ead{xavier.serra@upf.edu}

\cortext[cor1]{Corresponding author.}
\address[label1]{Universitat Pompeu Fabra, Barcelona, Spain}
\address[label2]{Sogang University, Seoul, Republic of Korea}

\begin{abstract}
Predicting the difficulty of playing a musical score is essential for structuring and exploring score collections. Despite its importance for music education, the automatic difficulty classification of piano scores is not yet solved, mainly due to the lack of annotated data and the subjectiveness of the annotations. This paper aims to advance the state-of-the-art in score difficulty classification with two major contributions. To address the lack of data, we present \emph{Can I Play It? (CIPI)} dataset, a machine-readable piano score dataset with difficulty annotations obtained from the renowned classical music publisher Henle Verlag. 
The dataset is created by matching public domain scores with difficulty labels from Henle Verlag, then reviewed and corrected by an expert pianist.
As a second contribution, we explore various input representations from score information to pre-trained ML models for piano fingering and expressiveness inspired by the musicology definition of performance. We show that combining the outputs of multiple classifiers performs better than the classifiers on their own, pointing to the fact that the representations capture different aspects of difficulty. In addition, we conduct numerous experiments that lay a foundation for score difficulty classification and create a basis for future research. Our best-performing model reports a 39.5\% balanced accuracy and 1.1 median square error across the nine difficulty levels proposed in this study. 
Code, dataset, and models are made available for reproducibility.
\end{abstract}


\end{frontmatter}

\input{1_introduction}

\input{2_relation}

\input{3_dataset}

\input{4_dataset_description}
\input{5_methodology}

\input{6_experiments}

\input{6b_case_study}

\input{7_conclusion}

\section*{Acknowledgements}
The authors would like to thank Craig Sapp, Luca Chiantore, and Pedro d'Avila for their
insightful comments. 

This work is supported in part by the project Musical AI - PID2019-
111403GB-I00/AEI/10.13039/501100011033 funded by the Spanish
Ministerio de Ciencia, Innovacion y Universidades (MCIU) and
the Agencia Estatal de Investigacion (AEI) and Sogang University Research Grant of 202110035.01.
\bibliography{template.bib}

\end{document}

%% file: 1_introduction.tex
\section{Introduction}
\label{sec:introduction}

Music corpora classification is a well-studied topic in Music Information Retrieval (MIR). It is frequently addressed from the perspective of listeners who explore, find, and receive song recommendations based on a search term, listening profile, or their search history \citep{ghosal2018music,weiss2015tonal,fukayama2016music,yang2010ranking}. 
However, there is a need to reframe this topic from the artist's perspective \citep{ferraro2021fair}. As artists often browse sound or score collections for creative or educational reasons, ongoing advancements in research related to this area \citep{lerch2019music} and the growing popularity of music education technologies \citep{hyon2022,eremenko2020performance,can2022violin} could potentially enhance their effectiveness in the future.

In this paper, we address the task of music score classification concerning performance difficulty,
a challenging and subjective task that remains largely unsolved \citep{chiu2012study,sebastien2012score,ramoneda2022}. More precisely, we are trying to answer the pianists' question when browsing an extensive collection of musical scores: ``Can I play it?''. Estimating the performance difficulty of the pieces is beneficial from music education and MIR perspectives.
Firstly, current piano performance repertoires comprise a limited number of musical works \citep{karlsen2015music}. The preference of institutions and teachers for popular or familiar pieces leads to a lesser focus on composers who are not as well-known. This manifests the long-tail problem, as described by \cite{levy2010music}. Simultaneously, the lack of tools for exploring extensive piano score collections  also dampens teaching and performance curricula diversity. These facts lead to several groups of composers, such as women and Eastern European composers, being historically under-represented and underplayed. 
Furthermore, involving students in creating their curriculum can increase their motivation, as they may not know their ability to play a music piece.

We pose the score difficulty classification task as an ordinal classification machine learning problem where the observations are machine-readable scores in MusicXML format and categories are numeric performance difficulty levels. Yet, a significant challenge in this endeavor is the data itself. Piano difficulty classification datasets have been notably restricted regarding availability, consistency, and scope.
Prior efforts varied from community-driven scores \citep{sebastien2012score,chiu2012study} to niche datasets like \emph{Mikrokosmos-difficulty}~\citep{ramoneda2022}, limited by its unique composition style. Addressing these gaps, we introduce the \emph{Can I play it? (CIPI)}~\footnote{Dataset available at:  \href{https://doi.org/10.5281/zenodo.6564421}{https://doi.org/10.5281/zenodo.6564421}} dataset, curated by matching metadata from various sources and difficulty annotations of the established editor Henle Verlag with the validation of expert pianists. CIPI features 652 classical piano pieces from 29 composers, spanning 9 difficulty levels, introducing the largest and most diverse open dataset with detailed difficulty annotations.

\begin{figure}[h!]
  \centering
  \includegraphics[width=1\textwidth]{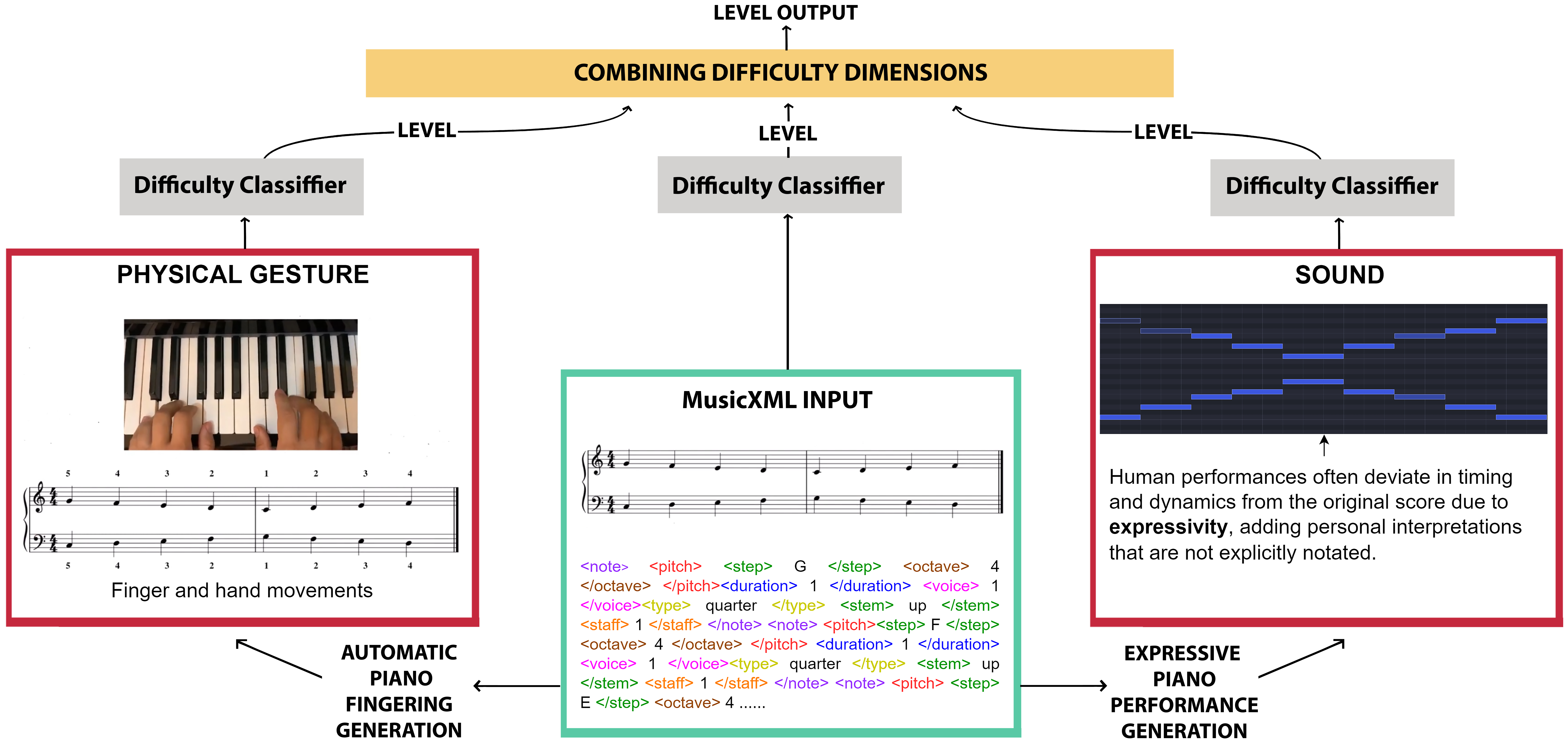}
    \caption{Diagram illustrating the flow of our methodology: From the digital score representation, both the physical gesture and sound are generated, simulating Cook's dimensions~\citep{cook1999analysing}. Consequently, the predictions from models trained on each representation are combined to estimate the difficulty of the original digital score.}
    \label{fig:venn}
\end{figure}

Based on musicological literature, our machine learning methodology aims to encapsulate the music performance's complexity. Even though scores provide a musically rich representation, the essence of music performance extends beyond mere notation, encompassing dimensions like physical gestures, sound, and verbal traditions, as described by \cite{cook1999analysing}. Emphasizing this multidimensionality, in this work, we research integrating the diverse elements of music performance: digital scores (notation), physical gestures, and sound. Guided by Cook's framework \cite{cook1999analysing}, we leverage interconnected concepts (depicted in Figure~\ref{fig:venn}) to potentially augment digital score data (notation) using pre-trained embeddings centered on physical gestures and sound and combine them.

Compared to our previous work \citep{ramoneda2022}, we now utilize a much larger and diversified dataset for difficulty classification. While the earlier method was mainly focused on physical gestures, our approach combines digital score data with physical and sound dimensions. To that extent, we introduced a new representation for the sound dimension of performance and refined the gesture representation. Using insights from previous research, notably \cite{jeong2019virtuosonet} on musical expressivity and \cite{ramoneda2022automatic} on finger and hand movements, we adapted models originally designed for generative tasks to our current predictive purpose. By leveraging their mid-layer embeddings, we predict difficulty in a manner consistent with \cite{cook1999analysing}'s performance definition.

\noindent The contributions of this article can be summarised as follows: 
\begin{itemize}[itemsep=0pt]

\item We introduce the \emph{CIPI} dataset, comprising 652 classical piano compositions that span 9 difficulty levels and represent 29 composers from the Baroque era to the 20th Century.

\item We train models on a range of musical performance dimensions, as inspired by \cite{cook1999analysing}. Further, by combining predictions from these models, we achieve a 39.5\% balanced accuracy and a median square error of 1.1 across the \emph{CIPI} dataset's 9 difficulty levels, surpassing single-model results.

\item We frame the score difficulty classification task as ordinal classification. Consequently, we evaluate the importance of choosing an appropriate loss function. Additionally, we introduce four evaluative metrics that encapsulate both the classification and regression dimensions of the problem.

\item  We design an architecture specifically for datasets with scores of varying lengths and piece-level difficulty annotations. Given the inherent challenge of different segments within a piece potentially having varied difficulty levels, termed weak labels, our approach incorporates context attention. This method optimizes the aggregation of patterns analyzed by the recurrent neural network, in contrast to previous approaches.

\item Our research also extends into detailed experimental procedures, paving the way for future studies. These include training strategies using partial compositions, exploring diverse techniques for feature fusion, applying a simplified classification system with only three categories on the \emph{CIPI} dataset, or focusing exclusively on the shorter musical pieces.

\item We provide the code, models, and dataset as open source to promote the research on the topic~\footnote{Code and models at: \href{https://github.com/PRamoneda/difficulty-prediction-CIPI}{https://github.com/PRamoneda/difficulty-prediction-CIPI} \\ Running code capsule available at:
 \href{https://codeocean.com/capsule/9645964/}{https://codeocean.com/capsule/9645964/}}. 

\end{itemize}

The remainder of this paper is organized as follows: in Section~\ref{sec:relation}, we describe the relation with previous work. In Section~\ref{sec:dataset}, we introduce the \emph{CIPI} dataset. Consequently, in Section~\ref{sec:methods}, we propose our difficulty classification methods. We present the results in Section~\ref{sec:experiments} and discuss some examples in Section~\ref{exp:examples}. Finally, in Section~\ref{sec:challenges},  we highlight potential avenues for future research, and in Section~\ref{sec:conclusion}, we state the main takeaways from the present research.

%% file: 2_relation.tex
\section{Relation with previous work}
\label{sec:relation}

In this section, we refer to the main computational methods to model difficulty in piano repertoire~\citep{sebastien2012score,chiu2012study,nakamura2014merged}. \cite{sebastien2012score} propose a list of different descriptors for difficulty classification, further extended by \cite{chiu2012study}. Consequently, \cite{nakamura2014merged,Nakamura2020} propose measuring the concept of difficulty based on automatic piano fingering models. Finally, in \cite{ramoneda2022}, we collect a small dataset, \emph{Mikrokosmos-dataset (MKD)}, and use neural networks based on score information and finger representations.

In \cite{sebastien2012score}, one of the initial works on score difficulty analysis, the authors propose seven descriptors to characterize the difficulty of piano scores. The authors use principal component analysis to project the descriptors onto two axes and evaluate the system's performance through expert perception. The list of descriptors is further extended by ~\cite{chiu2012study} up to 17 features for characterizing scores difficulty. 
However, neither of these proposals approaches the task as an ordinal classification, e.g., the predicted classes are ordered on an increasing difficulty axis. One of the limitations of the feature engineering methods is aligning with abstract notions very established in music, such as expressiveness or physical gesture. For instance, a study~\citep{chiu2012study} found the average pitch, calculated as the average of all the notes' pitches, relevant in predicting difficulty. However, it is not clear how this average pitch relates to the difficulty concept.
In addition, most of the features proposed by \cite{sebastien2012score,chiu2012study} are instrument agnostic and focus on score information. We argue that score difficulty analysis is a complex process dependent on the specific instrument and how the musician performs the music score~\citep{cook1999analysing}. To overcome this limitation, our approach aims to design instrument-specific methodologies and work with representations derived from performance rather than just musical structure. 
Finally, neither of the approaches open source their code, dataset, or models or provide sufficient information on creating the features, making replication difficult.

\cite{nakamura2014merged,Nakamura2020} propose estimating difficulty based on the fingering frequencies and playing rate. The rationale for this idea is that piano fingerings, which occur less often, increase difficulty. However, the proposal is not formally evaluated in the context of difficulty because of the lack of data available. This method is extended to other tasks such as polyphonic transcription~\citep{nakamura2018towards}, rhythm transcription~\citep{nakamura2017rhythm} or score reductions~\citep{nakamura2015automatic,nakamura2018statistical}, making clear the importance of piano technique in the creation of music technology systems.

In our previous work~\citep{ramoneda2022}, we propose a score difficulty classification method and introduce hand-crafted piano technique feature representations based on different piano fingering algorithms \citep{pianoplayer,nakamura2014merged} research. We use these features as input for two classifiers: a Gated Recurrent Unit neural network (GRU) with an attention mechanism and gradient-boosted trees trained on score segments. Our results show that incorporating fingering-based features into the score difficulty estimation task can improve performance compared to using only note-based features. This highlights the importance of considering the physical demands of playing the score when evaluating difficulty. Furthermore, the results demonstrate the potential of our proposed dataset, \emph{MKD}, for evaluating and comparing different score difficulty estimation approaches. Although it provides a valuable benchmark for further research in this area and offers insights into the influence of fingering on the perceived difficulty of piano scores, we think the dataset is small and focuses on one composer's work. We need to provide a more extensive and diverse dataset to impact the music community significantly.

%% file: 3_dataset.tex
\section{Can I play it? (\emph{CIPI}) dataset}
\label{sec:dataset}

We explored multiple sources comprising difficulty-labeled scores. However, not all sources were equally reliable, and not all contained public domain machine-readable scores. 
Publishers, examination boards, and online repositories often classify scores based on performance difficulty. However, the easiest levels frequently include 20th Century music scores without expired copyright.
Websites like 8note also distribute difficulty levels that users annotate. However, assessing the quality of these crowd-sourced annotations, recorded without any standard criteria, can be challenging.
After considering all alternatives, we selected Henle Verlag, a renowned publisher, as our source of difficulty labels. Their piano difficulty rankings are ranged from 1 to 9 and are annotated by Prof. Rolf Koenen~\citep{henle_notes}. Henle Verlag has an excellent reputation in the piano education community for producing high-quality and authoritative editions~\citep{jensen2020five,howat2013two}.

This section describes the methodology used to obtain difficulty labels for 2830 piano works from the Henle Verlag website.
Following open science practices, we aim to release the dataset and pair the difficulty labels with music scores from the public domain.
Note that Henle editions cover numerous pieces, but public domain scores may not be available for many of them. Finally, we discuss statistics about the dataset.

\subsection{Dataset creation methodology}
\label{sec:dataset_creation}

As a first step, we automatically match composer names and work titles
of Henle work to composer names and titles of pieces openly shared
by users of the public domain sources of \citep{musescore}, Craig Sapp Humdrum collection~\citep{chopin,bachi,beet,haydn,mozart,scarlatti,devaney2015theme,wtc} and Mutopia Project~\citep{mutopia}.
We downloaded pre-collected metadata ~\citep{xmander_reference}, previously used by MIR community \citep{Edirisooriya2021}.
More than a million MuseScore, two thousand Craig Sapp, and one thousand Mutopia entries are considered.
We use the Lucene~\citep{lucene} indexing and search engine to facilitate matching. Since there are no strict rules for crowd-sourced musical work naming, to improve the results, we index the names and aliases from MusicBrainzDB~\citep{swartz2002musicbrainz} for each work and composer, in addition to the Henle given titles. For example, for \emph{Clair de lune D flat major, Suite Bergamasque, Claude Debussy} there are more than 500 translations or aliases~\citep{MB_example}. Consequently, we assign each of the 949 Musescore, 390 Craig Sapp titles, and 299 Mutopia files a title from Henle, such that each title has  at least one assignment (7 assignments per title on average, ordered by Lucene hit score).

The three crowd-sourced collections used are in machine-readable formats, allowing access to all the information engraved in the musical score without transcribing. Note that we will use musicXML as the standard representation for the final dataset because it is the most interoperable and widespread format. Musescore music scores are converted to musicXML format without errors. In contrast, Craig Sapp and Mutopia music scores are manually validated to guarantee the proper conversion to musicXML from Humdrum and Lylipond formats, respectively.
However, crowd-sourced data may produce an uncertain quality in the collected scores and inconsistent metadata. 
To that extent, automatic matching produces two types of problems: false positives (e.g., matching the metadata of a piece with the wrong score) and including scores of doubtful quality. 
The retrieval system produces false positives because (a) the metadata is inconsistent, (b) the score is adapted for a different instrument, such as an adaptation for two violins of a piano piece, or (c) the score is arranged by a third author and the difficulty annotation is not valid. 
A typical error emerges when the score is automatically generated from audio, MIDI, or PDF without manually correcting it or when solely a score fragment is available. Another typical error is produced where the Henle annotation is for the whole musical piece, but only a movement is retrieved or the opposite.
To account for these problems, as an additional step, a manual best candidate selection or title rejection was performed by a human expert, a classical pianist with more than 20 years of playing the piano, a professional degree, and teaching experience.

\begin{figure}[h!]
  \centering
  \includegraphics[width=0.482\textwidth]{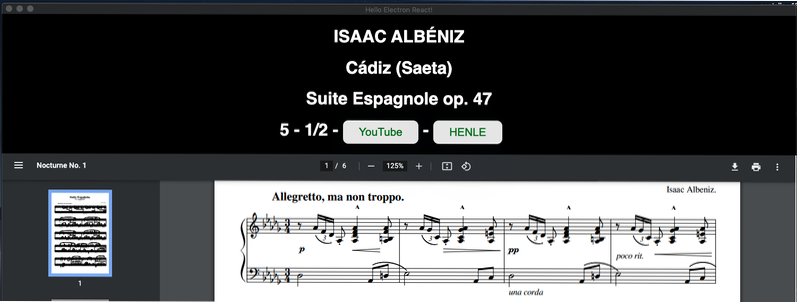}
    \caption{Annotation interface for reviewing and fixing automatic matches.}
    \label{fig:annotation_ui}
\end{figure}

To facilitate dataset validation and correction by the expert pianist, we create an electron~\citep{electron} desktop interface, shown in Figure~\ref{fig:annotation_ui}. This interface displays each piece, the associated metadata, and the annotated difficulty from the Henle Verlag Publisher. In addition, we include a link to the Henle website and a Youtube link we queried using the score metadata. The interface allows (i) to move across the retrieved musicXML scores with the horizontal computer keyboard arrows, and (ii) across the different pieces with the vertical arrows. It also allows (iii) to annotate the best option with the enter key, (iv) to indicate if any musicXML score corresponds to the Henle metadata, and (v) to display a confidence rating concerning the score quality on the scale of 1-4: (v.1) complete score, (v.2) a few minor engraving errors, (v.3) some signs of automatic creation and (v.4) not all the movements of the music work. The interface functionality allowed the expert pianist to manually correct and review all the pieces resulting from automatic matching in more than 100 hours of work and discard 58\% of the automated matches in the case of Musescore files, 42\% in the case of Craig Sapp matches, and 90\% in the case of Mutopia matches. Therefore, finally, the CIPI dataset comprises 394 Musescore files, 228 Craig Sapp files, and 30 Mutopia files, comprising 652 symbolic scores well engraved in musicXML format with annotations from the established Henle Verlag publisher.

\begin{table}[h!]
\centering
\begin{tabular}{l|l|l|}
\cline{2-3}
                                         & MKD & CIPI   \\ \hline
\multicolumn{1}{|l|}{No. composers}          & 1           & 29      \\ \hline
\multicolumn{1}{|l|}{No. levels}   & 3           & 9       \\ \hline
\multicolumn{1}{|l|}{No. pieces}    & 147         & 652 \\ \hline
\multicolumn{1}{|l|}{No. notes}    & 42699         & 1672699 \\ \hline
\multicolumn{1}{|l|}{No. measures} & 5041      & 115523  \\ \hline
\end{tabular}
\caption{Comparison between MKD and CIPI, showing the number of composers, levels, pieces, notes, and measures in both datasets.}
\label{comparison_mikrokosmos_difficulty}
\end{table}

\begin{figure}[h!]
  \centering
  \includegraphics[width=0.7\textwidth]{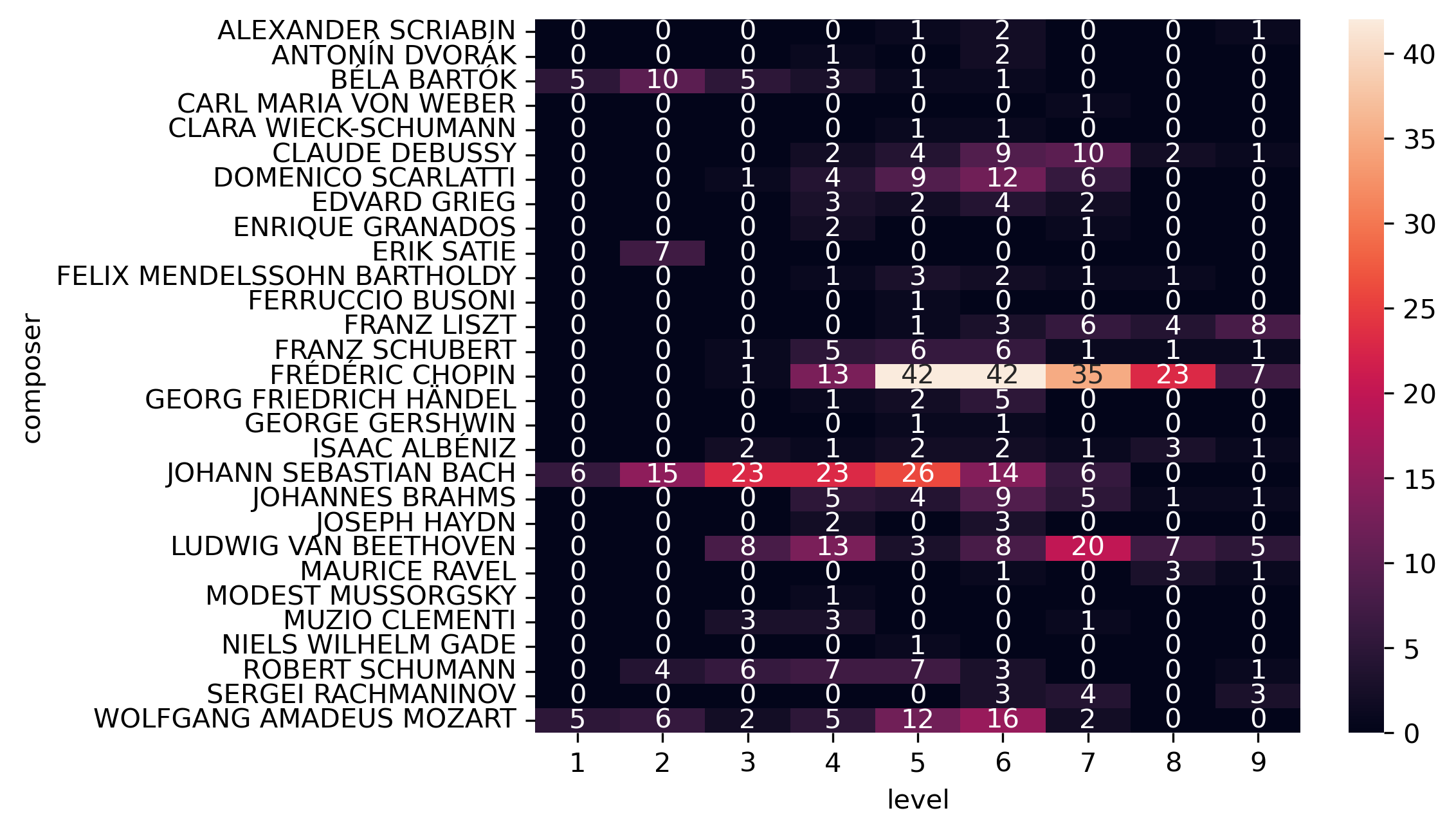}
    \caption{Heatmap displaying composers' distribution across the nine difficulty levels in the \emph{CIPI} dataset.}
    \label{fig:composer_heatmap}
\end{figure}

\begin{figure}[h!]
  \centering
  \includegraphics[width=0.7\textwidth]{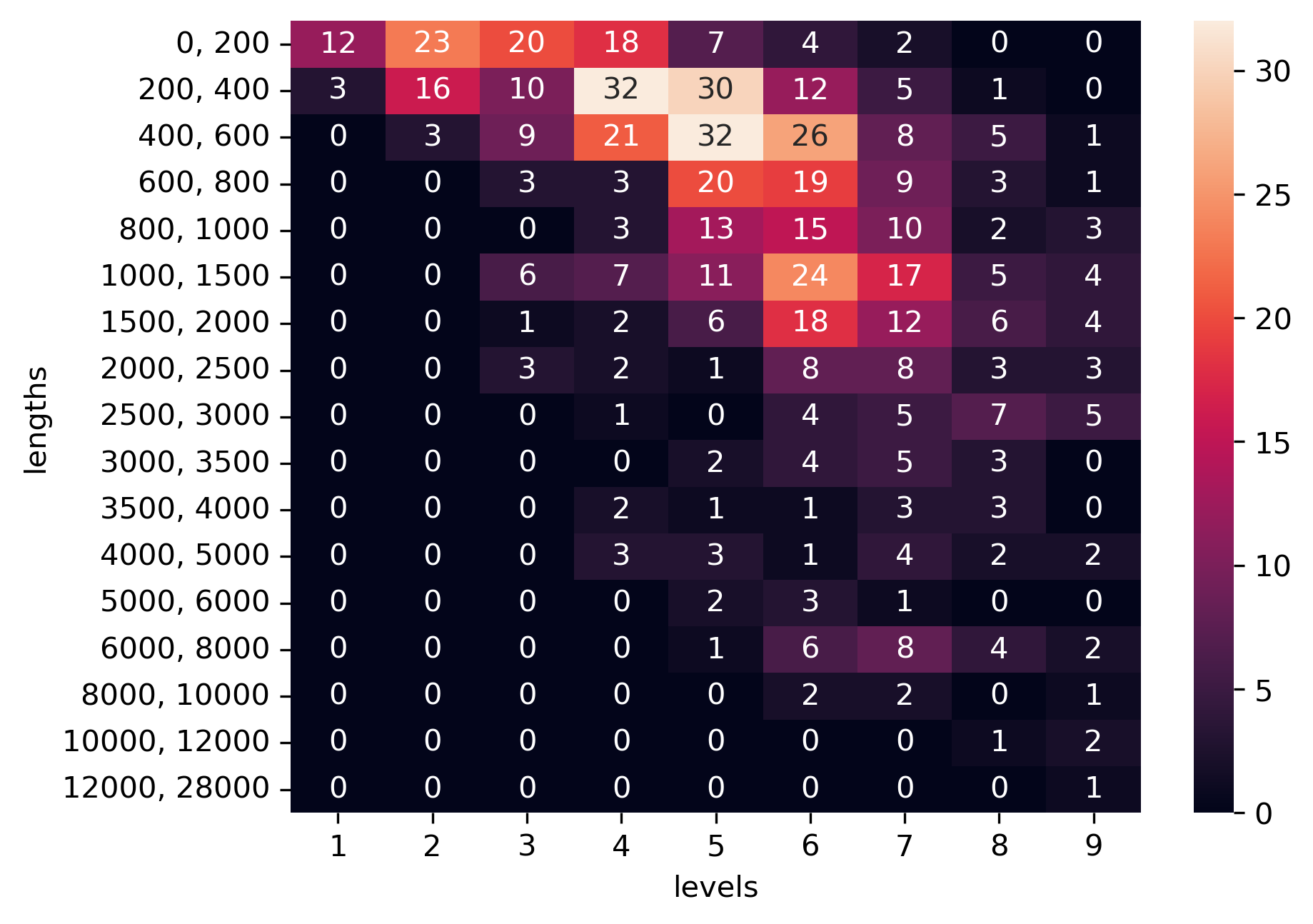}
    \caption{Heatmap displaying the distribution of the lengths (number of notes) across the nine levels of difficulty in the \emph{CIPI} dataset.}
    \label{fig:len_heatmap}
\end{figure}

\subsection{Dataset Analysis}

The \emph{CIPI} dataset comprises 652 classical piano pieces, spanning 9 difficulty levels and 29 composers ranging from the Baroque to the 20th Century. The distribution of composers and grades is shown in Figure~\ref{fig:composer_heatmap}. In comparison to the piano scores difficulty we previously released, \emph{MKD}, \emph{CIPI} is more diverse, as shown in Table~\ref{comparison_mikrokosmos_difficulty}. 

The \emph{CIPI} data distribution presents some challenges. Addressing the current biases by score engraving would involve significant expenses. However, with the rapid advancements in Optical Music Recognition \citep{calvo2020understanding}, such costs will soon become smaller. The distribution of the composers is skewed towards the most famous authors. The distribution is similar to other MIR datasets and real-word corpora~\citep{levy2010music} commonly known as long-tail. To that extent, some composers and styles are over-represented, as is shown in Figure~\ref{fig:composer_heatmap}. Note that we are using all the available scores at our disposal since the works of famous composers like Chopin were digitized more than those of less popular ones. These trends may bias the creation of personalized curricula based on the difficulty analysis research.
In addition, the grades have a bell-shaped distribution accumulating more in the central grade pieces than in the extremes. All these biases must be considered when using the CIPI Dataset.

The shorter pieces are over-represented in our dataset, while the longer pieces are few. We display a heatmap of the pieces and the corresponding difficulty levels in Figure~\ref{fig:len_heatmap}. Although the correlation between length and difficulty level is noticeable with Stuart's tau-c coefficient of $0.48$, the length is not the only feature important for characterizing difficulty level, e.g., some pieces with high difficulty levels are short. Therefore, we recommend in the future use of \emph{CIPI} dataset, pay special attention to biases related to the length of the pieces.

We distribute the \emph{CIPI} dataset for research purposes with the license creative commons 4.0, limiting access to the data upon request under the Zenodo platform. In addition, we distribute the links to all the source pieces, composer, and work metadata we have used in creating \emph{CIPI}.

%% file: 5_methodology.tex
\section{Methodology}
\label{sec:methods}

We introduce input representations based on the score information, expressive performance modeling of piano scores~\citep{jeong2019virtuosonet} and automatic piano fingering~\citep{ramoneda2022automatic}, as detailed in Section~\ref{ssec:backbone}. Furthermore, we employ a machine learning classification approach, discussed in Section~\ref{ssec:frontend}, to address automatic score difficulty classification on the \emph{CIPI} dataset. We also explore various methods for combining the musicology-inspired representations, as described in Sections~\ref{ssec:fusion} and \ref{ssec:ensemble}, and losses to capture the ordinal classification nature of the task.

\subsection{Backbone models}
\label{ssec:backbone}

Feature representations derived from inner-layer activations of pre-trained neural networks, commonly known as embeddings, may serve as powerful input features for downstream tasks~\citep{wang2019glue,raffel2020exploring,alonso2020deep}.
In section~\ref{sec:introduction}, we emphasize the importance of fingering and expressiveness features for capturing the information about piano performance~\citep{cook1999analysing}. Moreover, in section~\ref{sec:relation}, we discuss previous approaches that employ automatic piano fingering to indicate technical difficulty. In this context, we use the current state of the art in automatic piano fingering~\citep{ramoneda2022automatic} and the expressive piano performance generation model~\citep{jeong2019virtuosonet} as input features for neural networks to model piano difficulty. Both tasks are trained at the note level, allowing thousands of samples for each score in contrast with \emph{CIPI}, which has solely a global annotation for each score.


Each backbone model takes a sequence of notes-level encoding with $T$ notes $\textbf{N} = [n_1, n_2, \dots, n_T] $, which are extracted from MusicXML inputs. The features encoded in each note $n\in \mathbb{R}^d$ differ from each of the backbone models. Each backbone model calculates the learned hidden states of each note in $\textbf{N}$, which are then projected to output feature dimensions such as fingering probability distribution or expressive performance features.


\textbf{Automatic Piano Fingering - ArGNN backbone.} 
Predicting the physical hand and finger movements executed by a pianist based on a score could potentially serve as an indicator for assessing the difficulty of a piece~\citep{ramoneda2022,nakamura2014merged}. Fingering may be correlated with difficulty classifications since piano students learn to move their hands progressively during the early years of their careers while playing increasingly more challenging pieces.

The objective of piano fingering is to replicate a pianist's movement of the fingers on the piano keyboard while performing a particular piece of music.
It assigns a finger number to each note in the score from either the right or left hand: thumb (1), index (2), middle (3), ring (4), and pinky (5). According to \citep{palmer1997music}, piano fingering is among the most demanding human activities, usually requiring years of intensive practice. Piano players can improve their technique by adjusting their finger placement on the keys for adequate music interpretation. This involves anticipating the finger movements needed for the following sequence and adjusting accordingly, as the fingerings are not always clearly marked in the score.

We utilize a pre-trained auto-regressive graph neural network from our recent publication~\citep{ramoneda2022automatic} to produce embeddings that serve as input features for subsequent difficulty classification tasks. In this previous work, we train a model that predicts finger movements with near-human precision. The intermediate layers retain information about the input score and the predicted fingering movements, representing the music score (the music structure) and the physical movements associated with the technique. These are two main dimensions of music performance described in \cite{cook1999analysing}.

\begin{figure}[h!]
  \centering
  \includegraphics[width=0.5\textwidth]{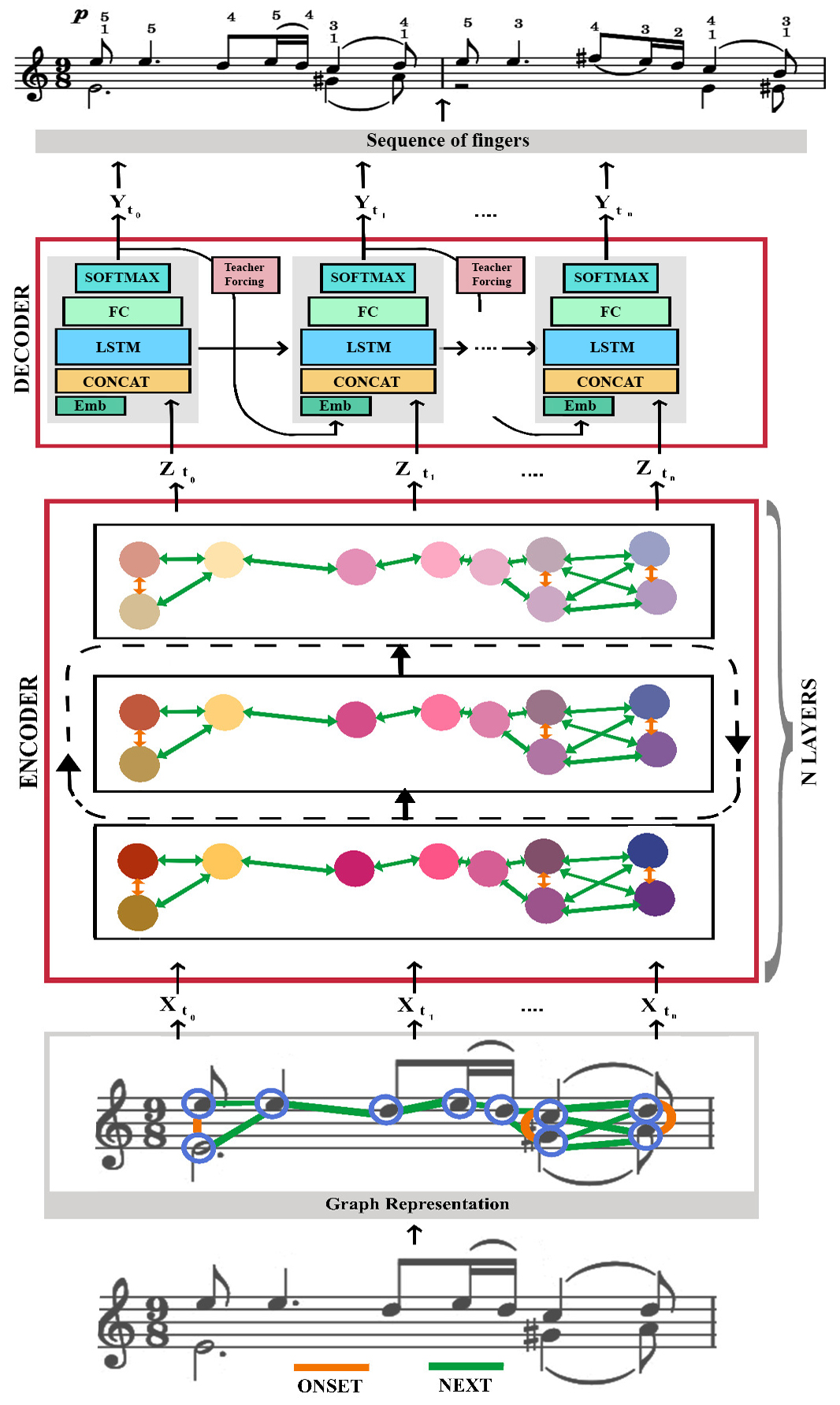}
    \caption{Encoder-decoder diagram of the autoregressive graph neural network for automatic piano fingering~\citep{ramoneda2022automatic} which we use as a proxy of \cite{cook1999analysing}'s technique dimension.}
    \label{fig:argnn_encoder_decoder}
\end{figure}

The ArGNN backbone has an encoder-decoder architecture, as shown in Figure~\ref{fig:argnn_encoder_decoder}. 
The encoder is a graph neural network (GNN), while an autoregressive recurrent neural network acts as the decoder. The input of the model is a sequence of notes containing information only about the pitch, similar to the previous literature on automatic piano fingering~\citep{nakamura2014merged,Nakamura2020,guan2022estimation}. The GNN encodes the polyphonic relation between notes, while the decoder ensures the sequential consistency of automatic piano fingering. We employ the last LSTM decoder layer embeddings as features for classifying scores based on performance difficulty. Due to its autoregressive nature, we decided to use this embedding because the intermediate representation in an autoregressive model includes information from preceding layers as well as previous temporal finger-label predictions.

\textbf{Expressive Performance Modelling.}
Musical expression and interpretation might be associated with difficulty. It requires the musician's understanding of the pieces to bring the emotional intent or meaning behind the music. These skills take time and practice to develop. As a result, music students gradually cultivate an understanding of music and its subsequent interpretations of dynamics and agogics, the so-called expressiveness.

We utilize the intermediate features from a neural network trained for expressive piano performance modeling, VirtuosoNet ~\citep{jeong2019virtuosonet}. 
To accurately estimate the performance features, the model must process the semantics of the music score, such as which note needs to be played with higher intensity or which part of a musical phrase, typically corresponding to the ending, should be played slowly. Therefore, the embeddings learned by the models trained for performance modeling may be used for other tasks, such as difficulty classification. 
As a result, we use the activations provided by one of the upper layers of VirtuosoNet as input features for another neural network that predicts difficulty.

VirtuosoNet takes a sequence of note-level score features and predicts note-level performance features. It consists of three modules: score encoder, performance encoder, and performance decoder. Here, we adapt the score encoder of a pre-trained VirtuosoNet to obtain note-level representations for the \emph{CIPI} Dataset. The score encoder includes specialized RNN layers for each musical hierarchy: note, voice, beat, and measure. The hidden representation of each RNN is broadcasted into note-level and then concatenated into note-level representations with 64 dimensions. 

\begin{figure}[h!]
  \centering
  \includegraphics[width=0.7\textwidth]{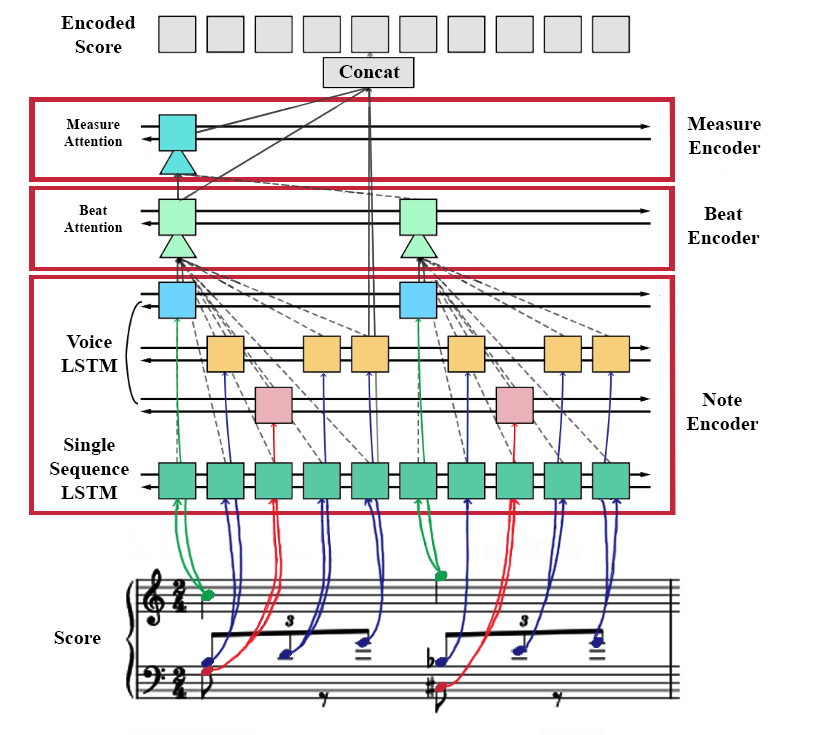}
    \caption{Hierarchical RNN-based diagram of the model for expressive piano performance generation~\citep{jeong2019virtuosonet} which we use as a proxy of \cite{cook1999analysing}'s expressiveness dimension.}
    \label{fig:virtuoso_encoder}
\end{figure}

\subsection{Classifier Architecture}
\label{ssec:frontend}

We propose using a straightforward architecture for summarising the performance difficulty of the musical pieces using the backbone features we describe in Section~\ref{ssec:backbone}. Similar architectures have been previously employed to benchmark and analyze proposed representations' language understanding \citep{wang2019glue}.

\begin{figure}[h!]
  \centering
  \includegraphics[width=0.6\textwidth]{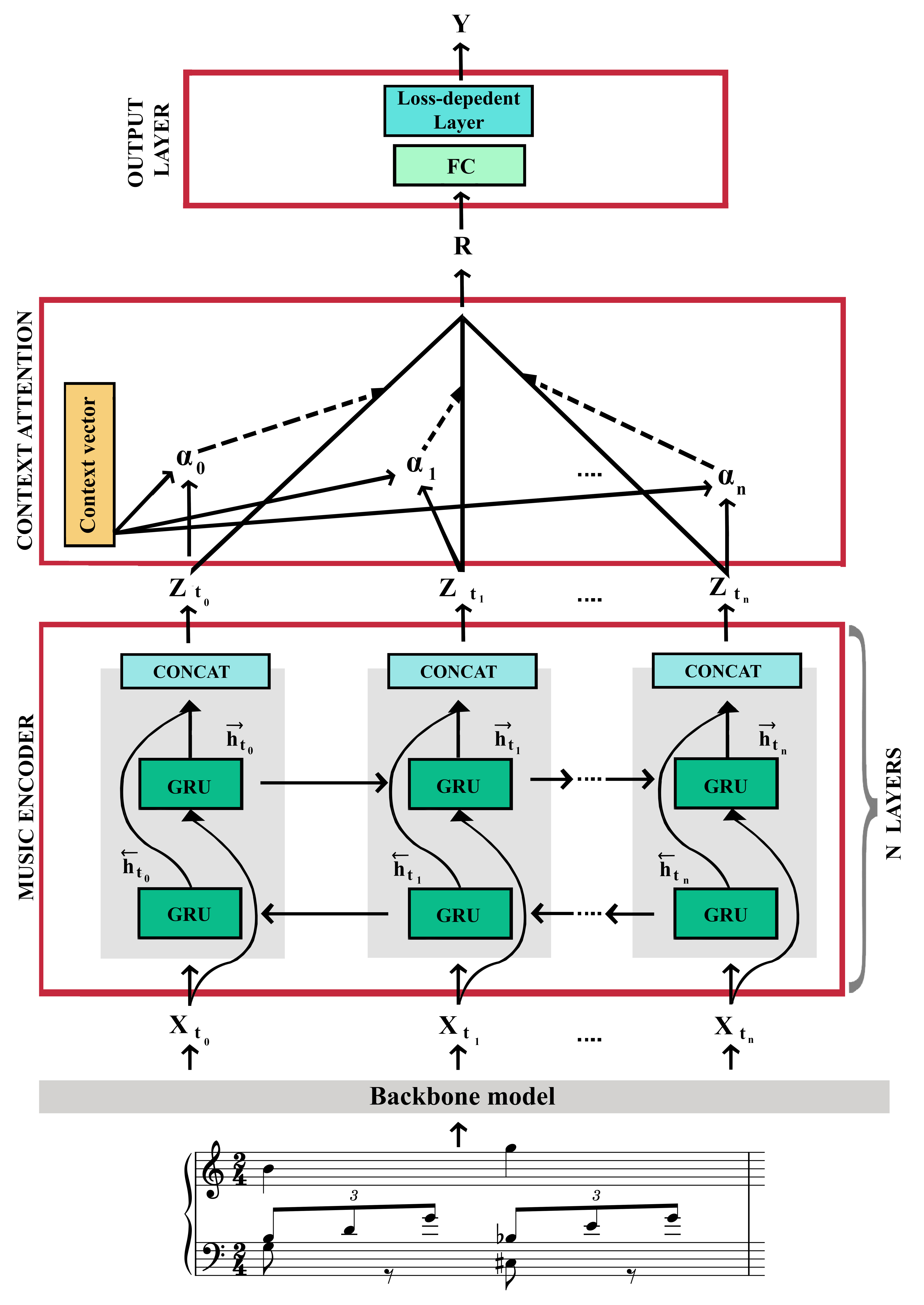}
    \caption{Diagram of the classifier architecture we use for score difficulty classification from the precomputed performance embeddings computed on the Backbone models.}
    \label{fig:attention_gru}
\end{figure}

In this work, we propose to change the previous architecture~\citep{ramoneda2022} by incorporating a different attention mechanism, context attention. It was first proposed for summarizing semantic meanings of a sentence or a paragraph for document classification~\citep{yang2016hierarchical} and was later adapted for modeling the hierarchical structure of music score in \cite{jeong2019virtuosonet}. In this work, we use hierarchical context attention to summarize the sequence of note-level hidden states of a piece with arbitrary length into a single vector, as shown in Figure~\ref{fig:attention_gru}.  

For a given sequence of note-level hidden states $\textbf{x}_T = [x_0, x_1, ..., x_t]$, hierarchical context attention summarizes it as a $y=\sum_t^T \alpha_t x_t$, where $\alpha_t=\text{Softmax}(\tanh(\textbf{W}x_t+b)^\top c)$ and $c$ represents a context vector that is trainable. In other words, the weight of each note representation is decided by the dot product value with the context vector. Since the context vector is a trainable parameter, context attention can learn which note is more important to predict the difficulty level of the piece. While the attention module of DeepGRU~\citep{maghoumi2019deepgru} uses the hidden state of the last time step to calculate attention weights, the context attention does not explicitly uses the last hidden state as a designated vector for attention calculation. Using the last hidden state can benefit gesture recognition as it was first proposed. Still, the difficulty of a musical piece does not need explicit focus on the beginning or the end of the piece. Therefore, we employed context attention instead of DeepGRU attention. The final layer, shown in Figure~\ref{fig:attention_gru}, is a linear layer (FC) followed by a loss-dependant layer discussed in Section~\ref{losses} to predict the difficulty level.

\begin{figure}[h!]
  \centering
  \includegraphics[width=0.8\textwidth]{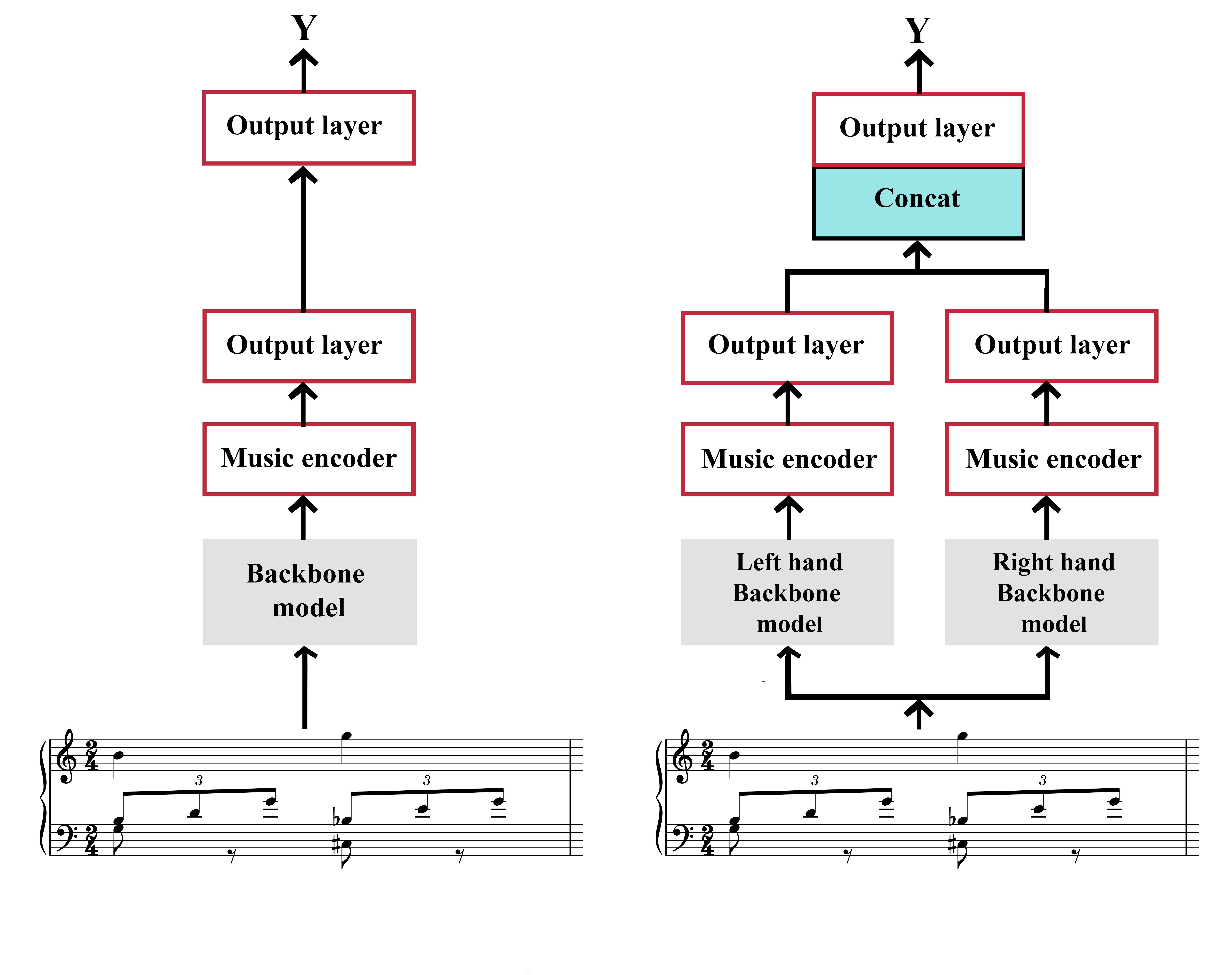}
    \caption{Comparison of classifier architectures showing (left) precomputed representations of \emph{virtuoso}, \emph{virtuoso\_enc}, or \emph{pitch} and a single branch classifier architecture, and (right) precomputed representation of \emph{argnn} and a classifier architecture with two branches.}
    \label{fig:two_branches}
\end{figure}

The automatic piano fingering backbone comprises two models, one for the right hand and the other for the left hand, which are trained independently. As the embeddings for each hand have different origins, we duplicate the GRUs and attention layers before the final output layer to accommodate each hand's different features, as shown in Figure~\ref{fig:two_branches}. 

\subsection{Feature Fusion}
\label{ssec:fusion}

We have proposed five strategies for combining the virtuoso, and ArGNN features to classify scores by performance difficulty. These strategies are based on early and late fusion approaches, which have been shown to improve different tasks~\citep{toselli2011multimodal,gao2020survey}, including music-related tasks~\citep{alfaro2022late}. Early fusion is applied as \emph{sync-fusion}, which concatenates the right and left-hand embeddings from the ArGNN representation with the virtuoso representation at each time frame. This technique only modifies the input to the classifier architecture.

On the other hand, late fusion methods modify the classifier architecture itself and include four different strategies: \emph{sum-fusion}, \emph{concat-fusion}, \emph{att-fusion}, and \emph{int-fusion}. The simplest of these, \emph{sum-fusion} and \emph{concat-fusion}, involve either summing or concatenating the outputs of the last layers from separate branches of the classifier, each dedicated to processing one of the input representations. The more complex methods, \emph{att-fusion} and \emph{int-fusion}, add a posterior architecture to the classifier to summarize the outputs of both branches using different attention mechanisms, providing a more sophisticated way of integrating information from the virtuoso and ArGNN embeddings.  The \emph{att-fusion} combines \emph{virtuoso} and \emph{argnn} branch outputs using the attention mechanism in Section~\ref{ssec:frontend}. \emph{int-fusion} uses the existing AutoInt~\citep{song2019autoint} feature fusion attention mechanism to combine the branches, automatically learning high-order feature interactions and mapping them into a low-dimensional space with a multi-head self-attentive neural network and residual connections.

These five fusion strategies showcase the flexibility of the proposed architecture in combining different aspects of musical performance to predict score difficulty. By considering both a piece's expressiveness and technical complexity, our model can provide a more comprehensive assessment of its difficulty, ultimately benefiting students, teachers, and performers in understanding and mastering the challenges presented by various compositions.

\subsection{Ensemble classifier}
\label{ssec:ensemble}

Ensemble methods typically improve the robustness and generalization of the classifier by reducing  overfitting and bias~\citep{opitz1999popular}. We propose to use a deep learning ensemble classifier to combine multiple models trained in different modalities to improve the classifier's overall performance. This is done by training multiple models on different representations on the same dataset and averaging their predictions to make the final decision. 

\subsection{Loss functions}
\label{losses}

The target variable we aim to predict represents increasing difficulty levels. Therefore, it is ordinal. However, standard classification algorithms do not consider the relationship of order between classes, which may yield inconsistent labels for the ordinal classes; e.g., consider a machine learning model that gives the probabilities values 0.5, 0.2, 0.9, which are inconsistent with consecutive difficulty levels 1, 2, and 3.  
Towards adapting the neural network to the ordinal nature of the problem we are trying to solve, we propose a wide range of solutions, from embedding ordinality into the loss functions to using regression instead of classification.

\textbf{The negative log-likelihood loss (NLLLoss)}. As a simple baseline, we use the NLLLoss, frequently applied to  multiple class classification. The last layer of the classifier architecture uses a logarithmic softmax function to output the probability distribution of the neural network with the size of the number of classes. The categorical index with a higher probability indicates the predicted class. Because our dataset is imbalanced, having more low-difficulty pieces, we use the weighted version of the loss by assigning a higher weight to less frequent difficulty levels.
The correct answer's probability value is added to the average after taking the log of the probability value following softmax. 
\begin{equation}
    \begin{split}
        \text{NLLLoss}(x, y) = -\frac{1}{N} \sum_{i=1}^{N} w_i y_i \log(x_i)
    \end{split}
\end{equation}
where $N$ is the number of samples in the dataset, $y_i$ is the label of the sample encoded as a one-hot vector, $x_i$ is the predicted probability of the sample belonging to each class, also encoded as a one-hot vector, and $w_i$ is a weighting factor for each sample, varying in function to the dataset imbalance.

\textbf{Mixed loss: regression and classification (RegClassLoss)}. We combine classification and regression losses to better model the data distribution and to avoid converging to sub-optimal minima. Towards taking advantage of the dual nature of difficulty score classification, we combine the NLLloss with a standard regression loss, in this case, the Mean Square Error (MSE) loss:

\begin{equation}
    \begin{split}
        \text{MSELoss}(l_{x}, l_{y}) = \frac{1}{N}\sum_{i=1}^{N} w_i(l_{yi} - l_{xi})^2
    \end{split}
\end{equation}
where $N$ is the number of samples, $l_x$ are the predictions and $l_y$ the ground-truth difficulty level as a scalar value, $l_{xi}$, $l_{yi}$ are, respectively, the predicted value and the true value of the $i-th$ sample, and $w_i$ is a weighting factor for each sample, varying in function to the dataset imbalance. Therefore, this loss minimizes the mean square error of an estimator, i.e., the square difference between the ground truth values and the estimated values.

To combine both losses, we add a projection layer that maps a scalar value from the same last hidden state of the classifier network. Therefore, the MSEloss uses a scalar, $l_{y}$, from the classifier's last layer as input, while NLLLoss uses $y$ as many scalars as classes. Finally, we combine both losses using a correction factor $\alpha$.

\begin{equation}
    \begin{split}
        \text{RegClassLoss}(x, l_{x}, y, l_{y}) = \text{NLLLoss}(x, y) + \alpha \cdot \text{MSELoss}(l_{x},  l_{y})
    \end{split}
\end{equation}

\textbf{Multilabel smoothed loss (MSLoss)}. We argue that difficulty is a subjective concept. It can change depending on how a piece is played and who plays it. Thus, we use a label smoothing on BCELoss, previously used to model subjectivity problems~\citep{lukasik2020does}. The BCELoss is usually applied to binary classification problems, with the predictions and the ground truth labels encoded as one-hot vectors.
To compute the smoothed labels, $\hat{y_i}$, we process the labels with a Gaussian smoothing function with $\sigma=0.5$ to train a multilabel prediction. 

We smooth the label one-hot vector using Gaussian blur to give slight weight to the neighboring difficulty level and a zero weight for the rest. This way, the model may account for difficulty-level subjectivity and produce more accurate predictions.
For example, a one-hot label $[0, 0, 0, 0, 1, 0, 0, 0, 0]$ is smoothed and normalized into $[0, 0, 0, 0.1, 1, 0.1, 0, 0, 0, 0]$.

\begin{equation}
    \begin{split}
        \text{MSmoothLoss}(x, y) =-\frac{1}{N} \sum_{i=1}^{N} w_i ( \hat{y_i} \log(x_i) + (1-\hat{y_i}) \log(1-x_i))
    \end{split}
\end{equation}
where $N$ is the number of samples in the dataset, $\hat{y_i}$ is the label smoothed, $x_i$ is the predicted probability of the sample belonging to each class, and $w_i$ is a weighting factor for each sample, varying in function to the dataset imbalance.

\begin{figure}[h!]
  \centering
  \includegraphics[width=0.5\textwidth]{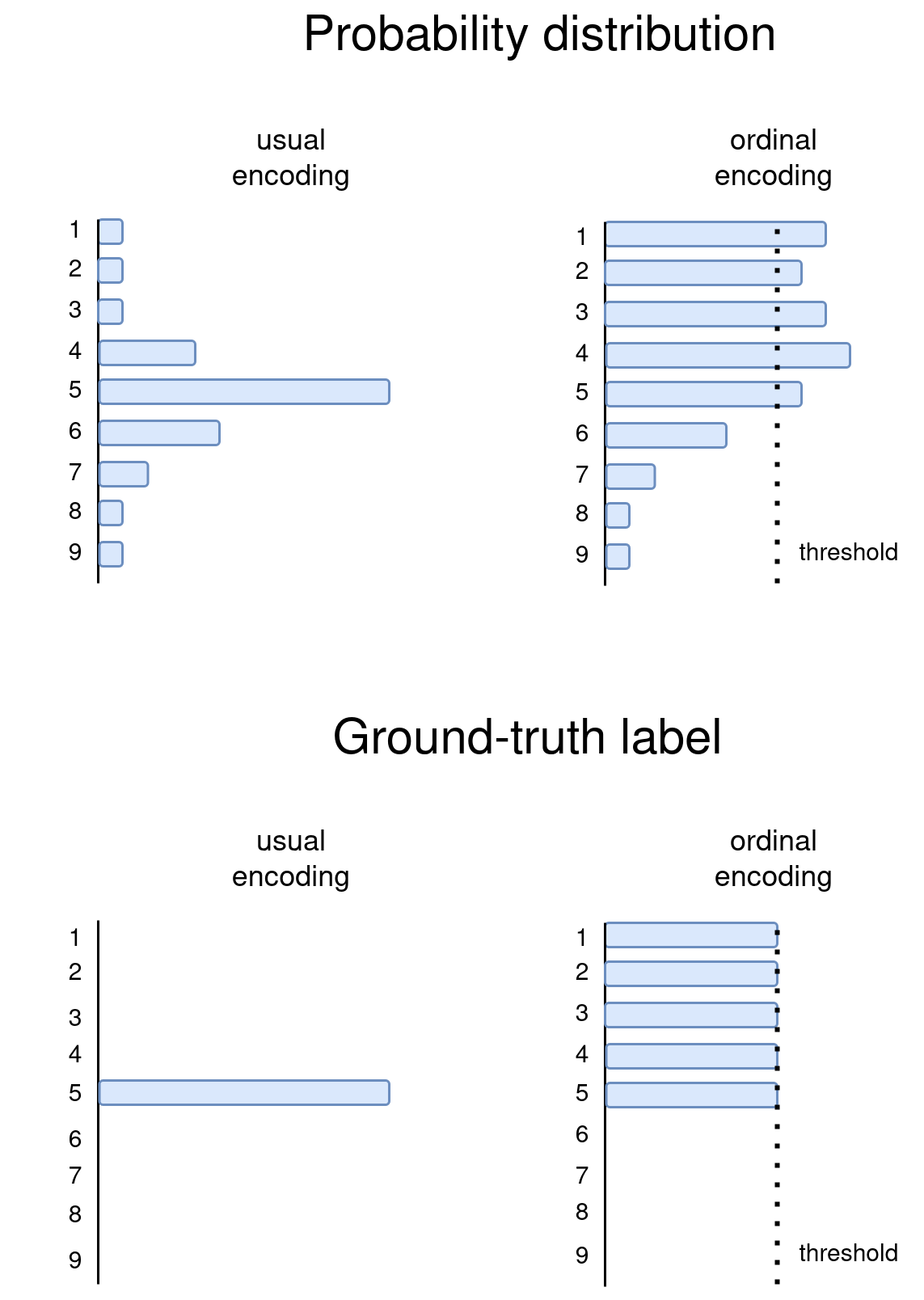}
    \caption{Comparison of the probability distribution and the ground-true label encoding for class level 5. (left) max likelihood encoding. (right) Ordinal encoding.}
    \label{fig:ordinal_encoding}
\end{figure}

\textbf{Ordinal loss (OrdinalLoss)}. The ordinal loss proposed by \citep{cheng2008neural} considers that the predicted labels have an ordinal relation between them. The proposal grounds in an ordinal encoding, shown in Figure~\ref{fig:ordinal_encoding}, in contrast to the one-hot encoding used in the previous approaches.

In ordinal encoding, the model is forced to learn an ordered structure where the prediction of one class implies that all the previous classes, following the order defined, are also predicted. Therefore, whether the model predicts a class with a higher encoded integer value, for example, class difficulty level 3, it also implies that classes: difficulty level 3, difficulty level 2 and difficulty level 2, and difficulty level 1 are also predicted, as shown in Figure~\ref{fig:ordinal_encoding}. To force the ordinal structure of the predictions, we use the mean squared error (MSE),

\begin{equation}
    \text{OrdinalLoss}(x, y) = \frac{1}{N}\sum_{i=1}^{N}(y_i - x_i)^2
\end{equation}

where $N$ is the number of samples in the dataset, $y_i$ is the ground-truth value of the sample, ordinal encoded, and $x_i$ is the prediction value of the sample. Both $y_i$ and $x_i$ are ordinally encoded and therefore have the same size as the number of classes. 
Note that in inference, a class is predicted if it reaches a certain threshold, in our case, $0.5$. Consequently, if the ordinal encoding structure is not satisfied in the prediction, not being continuous in terms of index that exceeds the threshold, the predicted class is not defined, and evaluation metrics compute it as an error. For example, if the prediction of the model is $[1, 0, 0, 0, 1, 0, 0, 0, 0]$ then the predicted label is not defined.

\textbf{Rank-consistent Ordinal Regression (CoralLoss).} The CoralLoss is a loss function for training neural networks for ranking tasks, introduced by \cite{cao2020rank}. It uses the same ground-truth encoding as the ordinal loss, shown in Figure~\ref{fig:ordinal_encoding}. The key distinction compared with OrdinalLoss is enforcing the model to learn a cumulative probability. This means that the model is required to learn a monotonic relationship between the class index and their probabilities, such that the logit for a class should always be higher than the logit of the preceding class.

As described in \cite{cao2020rank}, the weight parameters of the neural network, excluding the final layer's bias units, are denoted as $\textbf{W}$. The output of the penultimate layer, $g(\textbf{a}_i, \textbf{W})$, is shared by all nodes in the final output layer, with $K-1$ independent bias units added to $g(x_i,\textbf{W})$. The predicted empirical probability for class $k$ is given by $\widehat{P}(y_{i}^{(k)}=1)=\sigma(g(\textbf{a}_i,\textbf{W})+b_{k})$, where $\sigma(z)$ is the logistic sigmoid function.

The model is trained by minimizing the weighted cross-entropy loss function, which is defined as:


where $N$ is the number of samples, $\lambda^{(k)}$ denotes the weight of the loss associated with the classifier, and $K-1$ is the number of binary classifiers. Class 0 is implicitly encoded. During inference, the binary labels for rank prediction are obtained by $f_{k}(x_i)={\widehat{P}(y{i}^{(k)}=1)>\text{tr}}$ where tr is a fixed threshold.

%% file: 6_experiments.tex
\section{Experiments and results}
\label{sec:experiments}

We conducted a series of experiments to establish a foundation for performance difficulty classification and to understand better the relationship between input features, datasets, and losses in difficulty classification. Section~\ref{exp:setup} overviews the foundational experimental framework used for subsequent experiments. In Section~\ref{exp:basic}, we present the primary results. Section~\ref{exp:losses} evaluates the relationship between the input representation and losses proposed and Section~\ref{exp:combining} delves into the exploration of combining the multiple input representations. Finally, Section~\ref{exp:further} introduces additional experiments to understand better the task. 

\subsection{Experimental setup}
\label{exp:setup}

We base our experiments on two datasets containing scores and difficulty labels: the \emph{MKD} and the \emph{CIPI} dataset, which we detail in Section~\ref{sec:dataset}. Consequently, we evaluate our models using five pseudo-random splits to ensure reproducibility. For each split, we designate 60\% of the dataset for training, 20\% for validation, and reserve the remaining 20\% for testing. The results reported in the experiments represent the mean and standard deviation.

Our dataset stratification strategy considers both the target difficulty level and the length of the piece, which is discretized into corresponding intervals of 1000 notes. As a result, we partition the dataset into pairs consisting of the difficulty label and the corresponding length interval. It should be noted that specific pairs of difficulty labels and lengths are represented by fewer than five scores, which is less than the number of folds in our experimental setup. We randomly allocate these music scores to the training, validation, or testing sets.

In the \emph{CIPI} dataset experiments, multiple metrics are used to evaluate the difficulty classification task as an ordinal classification problem, including 9-class balanced accuracy Acc-9, \emph{3-class} balanced accuracy, Acc-3, relaxed class boundary accuracy, Acc±1, and mean square error, MSE. The dataset is highly unbalanced, so the metrics are macro-averaged to account for this. The Acc-3 metric groups the 9 levels into three groups of levels and computes balanced accuracy. The Acc±1 metric relaxes the class boundaries. Consequently, the predicted label mismatches with neighboring classes are not penalized. Finally, the MSE is used to analyze the regression potential of the task. The first two metrics, Acc-9 and Acc-3, are widely used classification metrics. In comparison, the latter two metrics, Acc±1 and MSE, are commonly used in regression and are related to the ordinal nature of the task.

The balanced accuracy, Acc-9 and Acc-3, is defined as,

\begin{equation}
    \text{Acc-n} = \frac{1}{n}\sum_{j=1}^{n} \frac{TP_j + TN_j}{TP_j + TN_j + FP_j + FN_j}
\end{equation}

Where $TP_j$, $TN_j$, $FP_j$, and $FN_i$ are the true positive, true negative, false positive, and false negative rates for the $j^{th}$ class, respectively, and $n$ is the total number of classes, i.e., 3 or 9.

The unbalanced mean square error is defined as: 

\begin{equation}
    \text{UMSE}= \frac{1}{N} \sum_{i=1}^{N} (y_i - \hat{y_i})^2
\end{equation}

Where $y_i$ is the ground-truth value for the $i^{th}$ sample, $\hat{y_i}$ is the predicted value for the $i^{th}$ sample, and $N$ is the total number of samples.

The unbalanced relaxed class boundary accuracy is defined as
\begin{equation}
     \text{UAcc±1} = \frac{1}{N} \sum_{i=1}^{N} \begin{cases} 1 & \text{if } | y_i - \hat{y_i} | \leq 1 \\ 0 & \text{otherwise}  \end{cases}
\end{equation}

Where $y_i$ is the ground-truth value for the $i^{th}$ sample, $\hat{y_i}$ is the predicted value for the $i^{th}$ sample, and $N$ is the total number of samples.
This way, it only penalizes when the predicted label mismatch is more significant than one. Otherwise, it is considered a correct classification.

We also macro-average both, UMSE and UAcc±1 metrics, for each class to consider the datasets' imbalance. Therefore,

\begin{equation}
    \text{MSE} = \frac{1}{n}\sum_{j=1}^{n} \text{UMSE}_j
\end{equation}

\begin{equation}
    \text{Acc±1} = \frac{1}{n}\sum_{j=1}^{n} \text{UAcc±1}_j
\end{equation}

Where $\text{UMSE}_j$ or $\text{UAcc±1}_j$ is the value of the metric, UMSE or UAcc±1, for the $j^{th}$ class and $n$ is the total number of classes.

Because the \emph{MKD} dataset contains solely 3 levels of difficulty, we evaluate the difficulty classification task with 3-class balanced accuracy. However, the annotations of the \emph{CIPI} dataset are structured on 9 levels. 
The \emph{MKD} dataset has only 3 classes and the regression errors are less critical, so we do not include the metrics Acc-3 and Acc±1 related to them.

The machine learning models are trained using early stopping. The metrics used on the validation set to determine the stopping point are Acc-9 and MSE. We use Acc-9 and MSE on the evaluation set to select the best-performing model in the early stopping. We train the deep learning methods using mini-batch stochastic gradient descent training using the Adam optimizer, a dropout of 0.2 between the GRU layers, gradient clipping with the value of $1\cdot10^{-4}$, a batch size of 64, and a learning rate of $1\cdot10^{-4}$. We use a balanced sampler, which retrieves a uniform distribution of the samples and weights in the losses for each batch to solve the problem of unbalanced data on the CIPI dataset. In the experiments on the \emph{MKD} dataset, we use the NLLLoss criterion as we have fewer classes, and the ranking criterion may cause large errors.

\subsection{Primary Results}
\label{exp:basic}

In this section, we outline the primary results of our research. A summarized overview of our experimental findings on the \emph{CIPI} and \emph{MKD} datasets is presented in Table~\ref{tab:final_cipi}. This table showcases the results of the best-performing models for each representation, specifically: \emph{argnn}, \emph{virtuoso}, and \emph{pitch}, along with their ensemble results. Notably, the representation \emph{virtuoso\_enc} is omitted as detailed in Section~\ref{exp:losses}. Models trained on \emph{CIPI} employ the OrdinalLoss criterion, while those trained on \emph{MKD} utilize NLLLoss.

The ensemble classifier's results demonstrate how combining these representations reduces errors and outperforms all other metrics. On the \emph{MKD} dataset, the balanced accuracy for all classes is $76.1$, while on the \emph{CIPI} dataset, it is $39.5$. This shows that the ensemble model performs better than models trained on individual representations, indicating that the representations are complementary. The performance improvement is more significant on the \emph{CIPI} dataset.

Moreover, the ensemble classifier's ranking metrics on the \emph{CIPI} dataset - with Acc±1 = $87.27$ and MSE = $1.1(0.2)$ - are outstanding. These results are appropriate for applications requiring understanding difficulty as a regression task, such as exploring large music libraries or curriculum learning.

\begin{table}[h!]
\centering
\begin{tabular}{lllll:l}
\cline{2-6}
          & \multicolumn{4}{c:}{CIPI}                            & MKD    \\ \cline{2-6} 
          & Acc-9        & Acc-3        & Acc±1        & MSE        & Acc-3   \\ \hline
argnn    & 32.6(2.8) & 69.2(3.6)  & 71.(5.3)  & 2.1(0.2) & 75.3(6.1)  \\
virtuoso & 35.2(7.3) & 67.(3.1) & 73.6(3.9) & 2.1(0.2) & 65.7(7.8)  \\
pitch        & 32.2(5.9) & 67.9(4.1)  & 76.4(2.8)  & 1.9(0.2) & 74.2(9.2) \\
\hdashline
ensemble & \textbf{39.5(3.4)} & \textbf{71.3(3.2)}  & \textbf{87.3(2.2)} & \textbf{1.1(0.2)} &  \textbf{76.4(2.3)} \\\hline
\end{tabular}
\caption{Experiment comparison of all the input representations proposed and the ensemble on \emph{CIPI} and \emph{MKD}.}
\label{tab:final_cipi}
\end{table}

\subsection{Input Representation and Losses for performance difficulty prediction}
\label{exp:losses}

We investigated the effectiveness of various representations and losses through several experiments. These findings are detailed in Table~\ref{tab:cipi_losses}. For a deeper understanding of the representations and losses, refer to Sections~\ref{ssec:backbone} and~\ref{losses}, respectively. As the class count increases, task ranking becomes increasingly critical, emphasizing the importance of task selection. Consequently, Table~\ref{tab:cipi_losses} primarily displays the NLLLoss results for the \emph{MKD} dataset (comprising 3 classes) and offers a comprehensive analysis of losses for the \emph{CIPI} dataset.

Our experiments show how different representations and losses perform in detail. Notably, while the representations \emph{argnn}, \emph{pitch}, and \emph{virtuoso} exhibit potential, the efficacy of OrdinalLoss as a loss function stands out. These insights will shape our experimental approach and guide future research to develop more precise and improved models.

\textbf{Representation results.} The \emph{argnn} representation on \emph{CIPI} demonstrates considerable variation in results depending on the loss experiment in question. Notably, the outcome was notable when employed with MSEloss for the Acc-3 metric, registering at $70.9$ with a standard deviation, $\sigma$, of $3.9$. In contrast, CoralLoss overcame others across all metrics, with Acc-9's second-best result across all representation experiments being $34.5$, accompanied by a $\sigma$ value of $3.6$.

For the \emph{virtuoso} representation on \emph{CIPI}, its performance was commendable during the OrdinalLoss experiment, registering MSE and Acc±1 values at $2.1$ and $73.$, respectively. It is worth mentioning that Acc-9 exhibited the best performance with a score of $35.2$ but was characterized by a high standard deviation of $7.3$.

When the \emph{pitch} representation on \emph{CIPI} was combined with MSloss, it yielded noteworthy results, especially for the Acc-9 and Acc-3 metrics, which were recorded at $33.6$ and $69.6$, respectively. However, OrdinalLoss exceeded other metrics in the Acc±1 and MSE metrics. The former metric displayed an exemplary result of $74.9$ with $\sigma = 3.5$, while the latter consistently exhibited superior performance across all representations.

The experiment on \emph{CIPI} employing the \emph{virtuoso\_enc} representation did not fare as well as its peers. Notably, with OrdinalLoss, mainly predicted overlapping classes with a subpar Acc-9.

Regarding the \emph{MKD} dataset, \emph{argnn} have a better performance than \emph{pitch} by a margin of 1.1\% with both \emph{virtuoso} and \emph{virtuoso\_enc} lagging. Even though these results were influenced by high standard deviations, when compared with \emph{CIPI}, they provide insights into the technical (physical gesture) nature of the \emph{MKD}.

Interestingly, the \emph{virtuoso} embedding representation surpassed the input of its original model, the \emph{virtuoso\_enc} representation. Concurrently, the \emph{argnn} representation aligned closely with the original input of the ArGNN-model, which is the \emph{pitch} representation. Given the subpar performance of the \emph{virtuoso\_enc} representation, it will be omitted from subsequent experiments.

\begin{table}[h!]
\centering
\begin{tabular}{lllll:l}

\cline{2-6}
\multicolumn{1}{l|}{} & \multicolumn{4}{c}{CIPI} & \multicolumn{1}{:l}{MKD} \\
\hline
\multicolumn{1}{l|}{loss}         & Acc-9        & Acc-3         & Acc±1         & MSE         & Acc-3         \\ \hline
                                  & \multicolumn{5}{c}{argnn}                                              \\ \hline
\multicolumn{1}{l|}{MSLoss}       & 30.4(6.8) & \textbf{70.9(3.9)}  & 66.(3.3)  & 2.6(0.5)  &  -  \\
\multicolumn{1}{l|}{NLLLoss}      & 26.6(7.2) & 63.9(3.8)  & 63.7(6.8)  & 3.5(2.1)  &  75.3(6,1) \\
\multicolumn{1}{l|}{CoralLoss}    & \textbf{34.5(3.6)} & 69.6(5.8)  & 69.7(5.6)  & 3.50(1)  &  -  \\
\multicolumn{1}{l|}{RegClassLoss} & 25.1(3.4) & 60.4(4.3)  & 65.4(7.4)  & 4.2(1.8)  &  -  \\
\multicolumn{1}{l|}{OrdinalLoss}  & 32.7(2.9) & 69.2(3.6)  & \textbf{72.(5.3)}  & \textbf{2.1(0.2)}  & -  \\ \hline
                                  & \multicolumn{5}{c}{virtuoso}                                           \\ \hline
\multicolumn{1}{l|}{MSLoss}       & 30.7(6.0) & 64.6(5.4)  & 66.6(4.4)  & 3.2(0.5)  & -   \\
\multicolumn{1}{l|}{NLLLoss}      & 26.5(4.8) & 55.8(3.8)  & 55.(4.3)  & 6.5(1.)  &  65,7(7,8)  \\
\multicolumn{1}{l|}{CoralLoss}    & 27.2(1.6) & 64.6(3.2)  & 60.(3.7)  & 5.5(1.4)  & -   \\
\multicolumn{1}{l|}{RegClassLoss} & 30.2(3.3) & 56.9(3.5)  & 59.9(2.5)  & 5.5(0.7)  & -   \\
\multicolumn{1}{l|}{OrdinalLoss}  & \textbf{35.2(7.3)} & \textbf{67.0(3.1)}  & \textbf{73.6(3.9)}  & \textbf{2.1(0.2)}  & -  \\ \hline
                                  & \multicolumn{5}{c}{pitch}                                                  \\ \hline
\multicolumn{1}{l|}{MSLoss}       & \textbf{33.6(4.5)} & \textbf{69.6(3.2)}  & 69.8(4.3)  & 2.3(0.4)  & -  \\
\multicolumn{1}{l|}{NLLLoss}      & 27.4(7.1) & 62.9(2.9)  & 51.8(6.6)  & 3.9(0.7)  & 74.2(5.3) \\
\multicolumn{1}{l|}{CoralLoss}    & 30.(2.4) & 65.8(2.7)  & 61.7(3.9)  & 4.6(1.5)  & -   \\
\multicolumn{1}{l|}{RegClassLoss} & 33.5(4.8) & 62.2(5.1)  & 51.(10.2) & 4.1(1.)  & -   \\
\multicolumn{1}{l|}{OrdinalLoss}  & 32.2(6) & 67.9(4.1)  & \textbf{76.4(2.8)}  & \textbf{1.9(0.24)}  & - \\ \hline
                                  & \multicolumn{5}{c}{vitruoso\_enc}                                                  \\ \hline
\multicolumn{1}{l|}{MSLoss}       & \textbf{25.6(8.9)} & \textbf{57.5(5.2)}  & 60.2(9.6)  & 3.6(0.9)  & - \\
\multicolumn{1}{l|}{NLLLoss}      & 19.8(5.6) & 47.9(9.1)  & 19.4(10.3) & 13.2(5.1) &  61.5(9.3) \\
\multicolumn{1}{l|}{CoralLoss}    & 13.4(9.) & 38.7(13.4) & 42.(10.2) & 7.3(2.1)  & - \\
\multicolumn{1}{l|}{RegClassLoss} & 16.7(8.7) & 44.4(9.2) & 25.(15.6) & 12.1(5.6) & - \\
\multicolumn{1}{l|}{OrdinalLoss}  & 12.(0.3) & 38.(4.3)  & \textbf{63.3(6.1)}  & \textbf{2.8(0.7)}  & - \\ \hline
\end{tabular}
\caption{Comparison of experimental results of models trained on all input representations (\emph{argnn}, \emph{virtuoso}, \emph{virtuoso\_enc} and \emph{pitch}) and all proposed losses (MSLoss, NLLLoss, CoralLoss, RegClassLoss, and OrdinalLoss).}
\label{tab:cipi_losses}
\end{table}

\textbf{Performance across losses.} The loss choice becomes essential given \emph{CIPI}'s inherent complexity, with an extensive class spectrum of difficulty gradations. Section~\ref{losses} described several losses pertinent to regression or ordinal classification, and their application is detailed in Section~\ref{exp:setup}. 

OrdinalLoss emerged as the superior loss function, especially when assessing performance through Acc±1 and MSE metrics. In contrast, when evaluating through Acc-9 and Acc-3 metrics, discerning a dominant loss function becomes challenging. However, OrdinalLoss consistently showcased robust modeling for the ordinal classification problem across representations, presenting competitive results in classification metrics.

Models employing OrdinalLoss consistently recorded better MSE, outshining other experiments across all representations. In the Acc±1 metric, which is closely related to MSloss, OrdinalLoss models excelled, marking differences ranging from $2.2\%$ to $7\%$. In light of these observations, our subsequent experiments will pivot around OrdinalLoss.

To sum up, Table~\ref{tab:cipi_losses} emphasizes the importance of loss selection for distinguishing between pure classification and ordinal classification. NLLLoss struggles due to its inability to recognize task ordinality. On the other hand, MSLoss performs well with minor differences in Acc-3. RegClassLoss models do not perform well with \emph{argnn} and \emph{virtuoso} embeddings. Coralloss and MSLoss have mixed results across metrics and representations. Finally, OrdinalLoss consistently demonstrated strong performance in regression metrics across various representations, yielding competitive scores in classification metrics.

\subsection{Combining multiple representations of performance difficulty}
\label{exp:combining}

\begin{table}[h!]
\centering
\begin{tabular}{lllll:l}
\cline{2-6}
\multicolumn{1}{l}{} & \multicolumn{4}{c:}{CIPI} & \multicolumn{1}{c}{MKD} \\ 
\cline{2-6}
              & Acc-9        & Acc-3        & Acc±1        & MSE        &  Acc-3      \\ \hline
sync-fusion   & 30.6(4.5) & 66.4(5.2) & 70.9(6.7) & 2.1(0.2) & 73.4(12.5) \\
concat-fusion & 32.7(3.) & 68.4(5.2) & \textbf{75.3(2.4)} & \textbf{1.9(0.3)} & 70.2(7.3) \\
sum-fusion    & 30.9(5.4) & 65.(4.9) & 72.3(3.1) & 2.2(0.2) & 70.5(13.4) \\
att-fusion    & 27.5(4.8) & 64.7(4.7) & 71.5(4.5) & 2.2(0.4) & 65.(4.1) \\
int-fusion    & \textbf{34.5(4.9)} & \textbf{69.5(2.3)} & 74.5(3.8) & 1.9(0.3) & 66.1(4.3) \\ \hline
\end{tabular}
\caption{The comparison of experiments evaluates the results of models trained on different strategies of feature fusion: \emph{sync-fusion}, \emph{concat-fusion}, \emph{sum-fusion}, \emph{att-fusion} and \emph{int-fusion}.}
\label{tab:fusion}
\end{table}

We explored the integration of distinct dimensions associated with piano performance, as discussed in Sections \ref{ssec:fusion} and \ref{ssec:ensemble}. Focusing on the representations \emph{argnn}, \emph{pitch}, and \emph{virtuoso}, we aimed to understand their cumulative effect on evaluating performance difficulty. Our investigation is divided into two groups of experiments: feature fusion and ensemble classification.

\textbf{Feature fusion experiments.} Merging different features is crucial in multi-mode settings. We mix \emph{argnn}, \emph{pitch}, and \emph{virtuoso} representations using various methods to learn about piano performance. We tested \emph{sync-fusion}, \emph{late-fusion}, \emph{concat-fusion}, \emph{sum-fusion}, \emph{att-fusion}, and \emph{int-fusion}.

Table~\ref{tab:fusion} displays the results. In our experiments, feature fusion does not significantly improves over models trained on separate representations. The best results were achieved by \emph{concat-fusion} and \emph{int-fusion}. Limited data might limit the potential of this merging. The results suggest that each method approaches a similar solution.

Comparing the methods on \emph{CIPI} and \emph{MKD}, more straightforward methods work better on the smaller \emph{MKD} dataset. For the bigger \emph{CIPI} dataset, the more intricate method, \emph{int-fusion}, performed better.

\textbf{Ensemble classification experiments.} Instead of merging features with feature fusion techniques, we grouped predictions through an ensemble. We grouped models using \emph{argnn}, \emph{virtuoso}, and \emph{pitch}, trained with OrdinalLoss. Table~\ref{tab:ensemble} shows the results. 

Grouping improved results significantly, suggesting these features complement each other, with the ensemble experiment improving all the metrics, Acc-9, Acc-3, Acc±1, and MSE, compared with previous experiments more than 4\%, 2\%, 7\% and, 0.8 points. By analyzing all combinations of ensembling models, we found that using all three representations achieved the best performance. This suggests that a holistic approach, considering technique, expressiveness, and score, leads to the best outcomes, and confirms the \cite{cook1999analysing} musicology definition of music performance.

\begin{table}[h!]
\addtolength{\tabcolsep}{-.25em}
\centering
\begin{tabular}{cllll:l}
\cline{2-6}
\multicolumn{1}{l}{} & \multicolumn{4}{c:}{CIPI} & \multicolumn{1}{c}{MKD} \\ 
\hline
\multicolumn{1}{l}{rep combinations} & Acc-9       & Acc-3       & Acc±1      & MSE  & Acc-3    \\ \hline
argnn                               & 32.7(2.9) & 69.2(3.6)  & 72.0(5.3)  & 2.1(0.2) & 75.3(6.1) \\
virtuoso                            & 35.2(7.3) & 67.(3.1)  & 73.6(3.9)  & 2.1(0.2) & 65.7(7.8) \\
pitch                                   & 32.2(5.9) & 67.9(4.1)  & 76.4(2.8)  &   1.9(0.2) & 74.2(9.2)  \\
argnn and virtuoso                  & 35.5(6.7) & 67.(0.9) & 80.9(4.3)  & 1.5(0.3) & 68.2(3.9)\\
argnn and pitch                         & 33.3(4.4) & 68.5(4.7) & 78.4(2.8)  & 1.6(0.3) & 70.7(4.1)\\
virtuoso and pitch                      & 33.4(3.2) & 66.5(5.2) & 80.8(4.6) & 1.5(0.2) &  72.3(3.1)\\ \hdashline
ensemble               & \textbf{39.5(3.4)} & \textbf{71.3(3.2)}  & \textbf{87.3(2.2)} & \textbf{1.1(0.2)} & \textbf{76.4(2.3)}\\ \hline
\end{tabular}
\caption{Ensemble ablation study: the figure displays the individual models first, followed by the models grouped in pairs, and finally, the outcome of the ensemble.}
\label{tab:ensemble}
\end{table}

\subsection{Other experiments}
\label{exp:further}

In the following experiments, we show further results that may provide valuable insights for future research in performance difficulty analysis. First, Section~\ref{exp:prev_work} relates the proposed work with the previous. Second, in Section~\ref{exp:fragments}, we show the challenge of training on fragments of the pieces. Consequently, in Section~\ref{exp:3classes}, we train the models at a higher granularity on the CIPI dataset. In Section~\ref{exp:strat}, we investigate the influence of different ways of stratifying the dataset, and in Section~\ref{exp:lengths} on the influence of the length of the music works on the task.

\subsubsection{Relation of architecture with previous work}
\label{exp:prev_work}

\begin{table}[h!]
\centering
\begin{tabular}{lllll}
\cline{2-5}
               & velocity           & argnn              & virtuoso          & pitch              \\
\hline
Ours   & \textbf{74.2(9.4)} & \textbf{75.3(6.1)} &\textbf{ 65.7(7.8)}         & \textbf{74.2(5.3)}       \\
\hdashline
DeepGRU& 68.6(13.1)& 67.3(9.1)          & 61.5(8.3)         & 54.2(12.1)           \\
\hline
\end{tabular}
\caption{Acc-3 results for various input representations for our model and the DeepGRU architecture on the \emph{MKD} dataset.}
\label{tab:prev_MKD}
\end{table}

This experiment compared our new architecture from Section~\ref{sec:methods} with the previously used DeepGRU model~\citep{ramoneda2022}. Our model is an improved version of the DeepGRU, designed to understand musical performances better. In addition, the previous study only used the \emph{velocity} representation. But we added three more: \emph{argnn}, \emph{pitch}, and \emph{virtuoso\_enc}, described in \ref{ssec:backbone}. Finally, previous \emph{velocity} can not be processed for most music scores in the \emph{CIPI} dataset. Therefore, we conduct this experiment only for the \emph{Mikrokosmos-difficulty (MKD)} dataset.

Table~\ref{tab:prev_MKD} shows that with DeepGRU, the \emph{velocity} representation achieves the best score of $68.6\%$, following the previous work trend. The \emph{ArGNN} and \emph{virtuoso} representations also have a significant performance with 67.3 and $61.5\%$. But both scores are lower than what we obtained with the new model. Also, the previous model obtained a reduced performance with \emph{pitch} representations.

\subsubsection{Training in fragments of the pieces} 
\label{exp:fragments}

We investigate the feasibility of training a model on small fragments of music pieces instead of the entire pieces. We assume that not all fragments share the same difficulty level, and the provided annotation only corresponds to the overall difficulty. Consequently, we trained the models with the whole piece in previous experiments. To verify this, we split each piece into fragments of 256 notes, including both hands, with a 25\% overlap. We conduct the same experimental setup described in Section~\ref{exp:setup} using the best-performing representations: \emph{argnn}, \emph{virtuoso}, and \emph{pitch}.

If different local fragments have a difficulty similar to the whole piece, the performance of the classifier trained on fragments should be comparable to that of the model trained on the full-length pieces. The number of samples in the dataset increases significantly when the fragments are considered, from 660 samples to 12769 samples. Therefore, if our assumption is not valid, performance may increase significantly.

The results, shown in Figure~\ref{tab:chunks}, reveal that the classifier's performance is not comparable with the previous experiment on \emph{CIPI}, with a difference of more than $10$ points in Acc-9 and $1$ point in the MSE metric. The experiments in any of the representations reach 20\% of nine class accuracy, while the Acc-3 metric is also underperforming. This indicates that the difficulty level of each piece fragment does not necessarily correspond to the overall piece difficulty level.

\begin{table}[h!]
\centering
\begin{tabular}{llllll}
\cline{2-6}
         & Acc-9       & Acc-3       & Acc±1      & MSE        & loss        \\ \hline
argnn    & 15.3(1.4) & 42.4(2.5) & 65.9(0.4) & 2.6(0.1) & 20.5(0.2) \\
virtuoso & 23.4(2.7) & 51.8(2.1) & 68.7(3.4) & 2.3(0.3) & 20.5(1.6) \\
pitch        & 20.0(1.4) & 49.7(2.8) & 68.3(4.4) & 2.4(0.2) & 21.1(1.9) \\ \hline
\end{tabular}
\caption{Experiment outcome of the models trained in fragments of the pieces instead of using as input the whole piece on the \emph{CIPI} dataset. }
\label{tab:chunks}
\end{table}

\subsubsection{Training in three classes}
\label{exp:3classes}

In Section~\ref{exp:basic}, we found that the Acc-3 trained on the \emph{CIPI} dataset is comparable with \emph{MKD} Acc-3 performance. In this section, we explore on \emph{CIPI} dataset whether the performance of the model can be improved by training it on fewer classes. Specifically, we consider if training in only three classes can enhance the Acc-3 metric. The classes predicted will be divided into $class\ 1\ =\ {1,\ 2,\ 3}$, $class\ 2\ =\ {4,\ 5,\ 6}$ and $class\ 3\ =\ {7,\ 8,\ 9}$, i.e., the same division in which Acc-3 is evaluated. We compare the representation of better-performing representations: \emph{argnn}, \emph{pitch}, and \emph{virtuoso} while optimizing the NLLLoss because the ordinal ranking nature of the task is less critical when only three classes are predicted.

The results of training on only three classes are presented in Table~\ref{tab:3-classes}. It can be observed that the \emph{argnn} representation outperforms the \emph{pitch} and \emph{virtuoso} representations with more than 3\%. Furthermore, the models perform worse than the Acc-3 predicted in Table~\ref{tab:final_cipi}. Thus, it can be concluded that training on the nine classes, as shown in Table~\ref{tab:final_cipi}, results in better Acc-3 performance than only training on three classes. The results of this experiment have a reduced accuracy varying from $1.1$ in the case of \emph{argnn} to $7.0$ in the case o \emph{pitch}. In other words, it can be claimed that training in nine classes does not negatively impact the Acc-3 metric.

\begin{table}[h!]
\centering
\begin{tabular}{lll}
\cline{2-3}
         & Acc-3        & MSE       \\ \hline
argnn    & 68.1(6.1) & 0.3(0.1)                \\
pitch       & 60.9(4.6) & 0.37(0.1)               \\
virtuoso & 59.1(5.2) & 0.47(0.1)              \\ \hline
\end{tabular}
\caption{Experiment outcome of the models trained in three classes on the \emph{CIPI} dataset.}
\label{tab:3-classes}
\end{table}

\subsubsection{What is the effect of stratifying?}
\label{exp:strat}

\begin{table}[h!]
\centering
\begin{tabular}{lllll}
\cline{2-5}
         & Acc-9         & Acc-3        & MSE         & loss       \\ \hline
\multicolumn{5}{c}{stratify by length and difficulty level}                          \\ \hline
argnn                               & 32.7(2.9) & 69.2(3.6)  & 72.(5.3)  & 2.1(0.2)  \\
virtuoso                            & 35.2(7.3) & 67.(3.1)  & 73.6(3.9)  & 2.1(0.2)  \\
pitch                                   & 32.2(5.9) & 67.9(4.1)  & 76.4(2.8)  & 1.9(0.2)  \\ \hline
\multicolumn{5}{c}{stratify by composer and difficulty level}                        \\ \hline
argnn    & 29.7(6.7) & 65.8(4.5) & 68.6(5.4) & 2.4(0.2) \\
virtuoso & 26.6(7.4) & 65.5(6.) & 75.6(1.7) & 1.9(0.2) \\
pitch    & 26.9(1.1) & 65.2(4.9) & 74.8(3.2) & 2.(0.3) \\ \hline
\end{tabular}
\caption{Experiment outcome of the models trained with different an alternative stratification on the \emph{CIPI} dataset.}
\label{tab:strat}
\end{table}

In Section~\ref{exp:setup}, we decided to stratify the dataset by lengths and difficulty levels. In this experiment, we aim to compare this decision with other alternatives of stratifying. One main alternative is stratifying by difficulty levels and composer. We do not employ a combination of the two previous approaches for stratification because when considering different sets of difficulty levels, compositions, and lengths, more than half of the dataset cannot be stratified as there are less than five samples in those sets.

Table~\ref{tab:strat} compares the stratification of the dataset by length and difficulty levels versus that by composer and difficulty levels. More sensitivity can be observed when using the latter method. If we compare each model trained with a certain representational can keep the dropout of more than $3$ points in Acc-9. The lack of generalization may be because some subsets, such as the training, validation, or test sets, have sequences of a certain length grouped in a specific class. The model seems more sensitive to lengths compared to the composer. Further research is needed to understand this phenomenon.

\subsubsection{How much influence the lengths into the results?} 
\label{exp:lengths}

\begin{table}[h!]
\centering
\begin{tabular}{llllll}
\hline
lengths & Acc-9       & Acc-3       & Acc±1      & MSE        & epoch time \\ \hline
\multicolumn{6}{c}{argnn}                                                   \\ \hline
full    & 32.7(2.9) & 69.2(3.6) & 72.(5.3) & 2.1(0.2) & 11s        \\
7000    & 25.4(5.) & 62.0(6.4) & 73.7(2.8) & 2.(0.2) & 4s         \\
3500    & 21.9(4.2) & 57.4(6.1) & 74.5(4.8) & 2.2(0.4) & 2s         \\ \hline
\multicolumn{6}{c}{virtuoso}                                                \\ \hline
full    & 35.2(7.3) & 67.(3.1) & 73.6(3.9) & 2.1(0.2)  & 9s         \\
7000    & 32.1(7.5) & 62.8(7.3) & 69.6(4.3) & 2.3(0.3) & 4s         \\
3500    & 26.6(7.4) & 58.(6.)  & 72.3(4.)  & 3.2(0.4)  & 2s         \\ \hline
\multicolumn{6}{c}{pitch}                                                       \\ \hline
full    & 32.2(6.) & 67.9(4.1) & 76.4(2.8) & 1.9(0.2) & 10s        \\
7000    & 25.6(4.6) & 65.8(5.) & 73.2(3.3) & 2.(0.2) & 4s         \\
3500    & 28.1(5.7) & 59.5(5.) & 76.6(3.) & 1.9(0.3) & 2s         \\ \hline
\end{tabular}
\caption{Experiment results of models trained on \emph{argnn}, \emph{virtuoso}, and \emph{pitch} on the \emph{CIPI} dataset for pieces of any length (full), pieces less than 7000 notes (7000), and pieces less than 3500 notes (3500).}
\label{tab:lengths}
\end{table}

We examine whether the variable length of music pieces in the \emph{CIPI} dataset affects performance. In Section~\ref{sec:dataset}, we illustrate the differences in length between pieces in \emph{CIPI} using Table~\ref{fig:len_heatmap}, which shows that some pieces have fewer than 500 notes while others have more than 30,000 notes. In the first experiment, we remove the pieces exceeding 3500 notes from the original splits, while in the second experiment, we remove the pieces exceeding 7000 notes. The splits were calculated by stratifying by composers and lengths, as outlined in Section~\ref{exp:setup}. Consequently, removing pieces greater than 3500 and 7000 notes, we preserve the same proportion of evaluation sets (train: 60\%, validation: 20\%, test: 20\%) and the stratification.  Therefore, we trained and evaluated the models over pieces of all possible lengths (\emph{full}), pieces with fewer than 7000 notes (7000), and pieces with fewer than 3500 notes (3500). We compare the top three representations: \emph{argnn}, \emph{virtuoso}, and \emph{pitch}, with OrdinalLoss.

The results are presented in Table~\ref{tab:lengths}. The classification metrics, Acc-9 and Acc-3, are lower in 7000 and 3500 experiments, with decrements ranging from 2\% to 11\%. In contrast, ordinal classification metrics, MSE and Acc±1, remain similar, with less than $0.2$ increment or decrement. Notably, the computational time cost is substantially reduced—close to 60\% in the 7000-note experiments and 80\% in the 3500-note experiments. These results offer valuable insights for future research, highlighting the trade-off between performance and piece length and underscoring potential avenues for speeding up computations in future studies.

%% file: 6b_case_study.tex
\section{Case study}
\label{exp:examples}

The case study explores particular examples and their difficulty predictions of the models trained on, \emph{argnn}, \emph{virtuoso}, and \emph{pitch}, and the final predictions of combining the three models, \emph{ensemble}. To carry out the case study, we checked the examples with a more significant typical deviation between the models trained on \emph{argnn}, \emph{virtuoso}, and \emph{pitch}, shown in Table~\ref{tab:final_cipi}. The study provides a comprehensive analysis of music scores, using multiple examples to demonstrate the performance of the models. The objective of the case study is to gain insights into how these models process and interpret musical information and how we can design accurate score difficulty classification systems based on combining the multiple music performance dimensions.

Furthermore, the study also identifies common errors detected during the analysis. These errors provide valuable information for future research and can help researchers improve the models' performance. The results of this case study can contribute to the advancement of music information retrieval and can significantly impact the development of more sophisticated music difficulty prediction models.

\begin{figure}[h!]
\centering
\subfloat[a]{
  \includegraphics[clip,width=0.6\columnwidth]{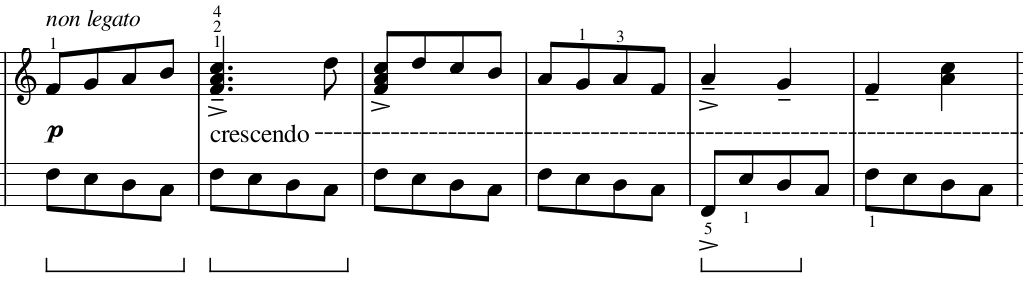}
}

\subfloat[b]{
  \includegraphics[clip,width=0.6\columnwidth]{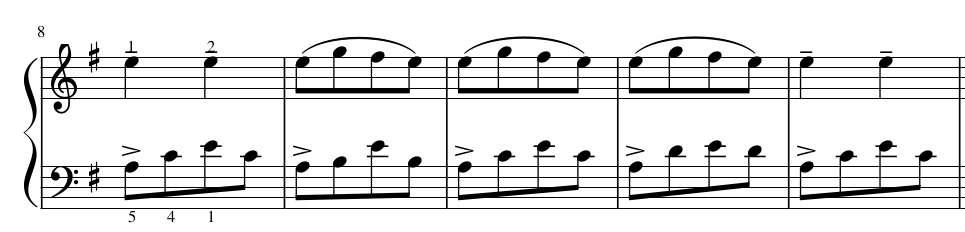}
}

\caption{Examples of pieces with the model trained in \emph{virtuoso} predicting overestimated difficulty. (a) Fragment of the piece \emph{Winter solstice song, Béla Bartók}. (b) Fragment of the piece \emph{Children's Dance, no. 10, Béla Bartók}.}

\end{figure}
\label{fig:case1}

We have occasionally observed how the \emph{pitch} and \emph{argnn} models have a lower prediction than the models trained on \emph{virtuoso}. We want to show it with examples in Figure~\ref{fig:case1}. The first example is 
\emph{Winter solstice song, Béla Bartók}, the prediction of the ensemble and the ground truth is level 3 while \emph{argnn} and \emph{x} prediction are 1 and \emph{virtuoso} prediction is 6. The second example is \emph{Children's Dance, no. 10, Béla Bartók}. The ground truth label is 2, the \emph{ensemble} prediction is 3, and the predictions of \emph{argnn} and \emph{x} are the same as the ground truth. In contrast, \emph{virtuoso} model outputs level 6. The piece lies in the constant changes in dynamics and articulation, even though the fingering is simple and there is limited use of cross-fingerings and note patterns. In addition, keeping the 4/4 time and the local tempo also requires skill and precision. However, although it is difficult to assess whether the \emph{argnn} or \emph{virtuoso} predictions are accurate, the score information, analyzed by \emph{pitch},  may be simpler.

\begin{figure}[h!]
\centering
\subfloat[][c]{
  \includegraphics[clip,width=0.7\columnwidth]{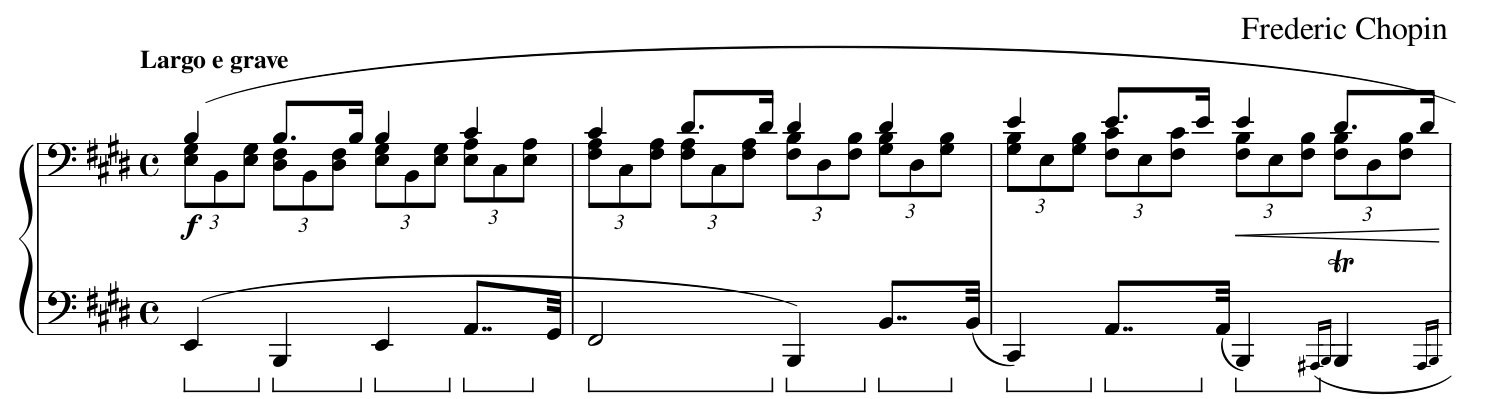}
}

\subfloat[][a]{
  \includegraphics[clip,width=0.7\columnwidth]{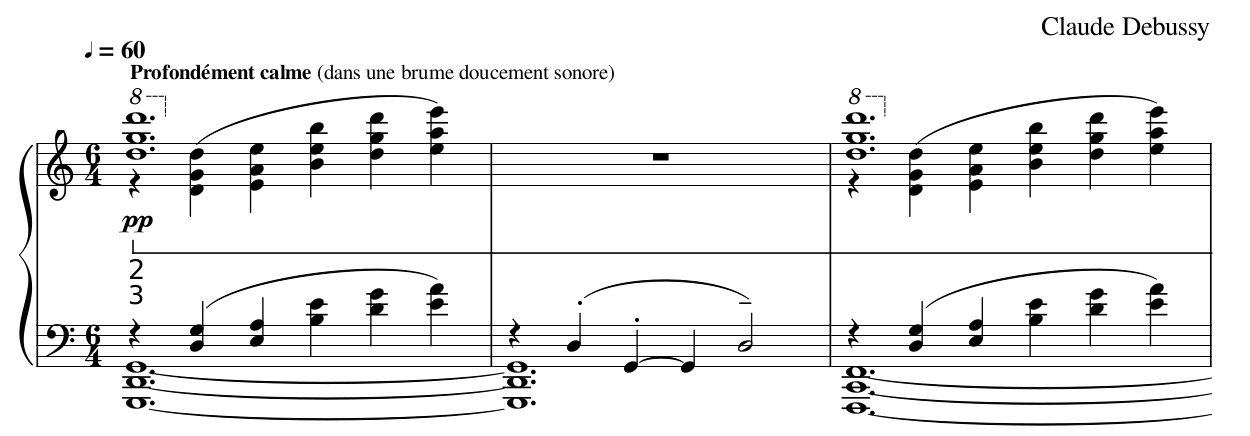}
}

\caption{Examples of pieces where the model trained in \emph{virtuoso} have a low estimated prediction. (a) Fragment of the piece \emph{Prelude E major op. 28,9, Frédéric Chopin}. (b) Fragment of the piece \emph{La Cathédrale engloutie, Preludes, Claude Debussy}.}
\label{fig:case2}
\end{figure}

In Figure~\ref{fig:case2}, we show two examples where the \emph{virtuoso} model estimates lower than the ensemble's other two models. The first piece is \emph{Prelude E major op. 28,9, Frédéric Chopin}. The ground truth label is 5, and the ensemble model's prediction is 5. The prediction made by the \emph{argnn} model is 8, while the \emph{pitch} model predicts 5 and the \emph{virtuoso} model predicts 2. 
The second piece is \emph{La Cathédrale engloutie, Preludes, Claude Debussy}. The ground truth label is 5, and the ensemble model's prediction is 8. The prediction made by the \emph{argnn} model is 9, while the \emph{pitch} model predicts 7 and the \emph{virtuoso} model predicts 3. Both pieces present unique and challenging finger sequences that can prove difficult for many pianists, and \emph{argnn} model may capture those patterns. However, it is uncertain if their technical demands are as simple as predicted, or if it may be a bias due to the pieces' relatively slow tempo.

\begin{figure}[h!]
\centering
\subfloat[][a]{%
  \includegraphics[clip,width=0.6\columnwidth]{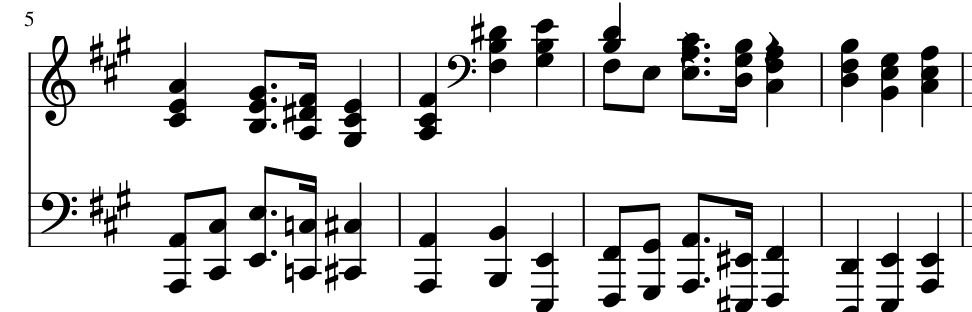}%
}

\caption{Example of under-performance of the model trained on \emph{virtuoso}. Fragment of \emph{Wichtige Begebenheit op. 15,6, Robert Schumann}.}

\label{fig:case3}
\end{figure}

We have observed some scores engraved without the articulations, as shown in Figure~\ref{fig:case3}. However, some editions or composers do not provide the articulations, and we think is important to expose this case. The piano piece is \emph{Wichtige Begebenheit op. 15,6, Robert Schumann}. The ground truth label is 4, and the ensemble model's prediction is 5. The prediction made by the \emph{argnn} model is 7, while the \emph{pitch} model predicts 5 and the \emph{virtuoso} model predicts 2. We can observe the large performance deterioration of the \emph{virtuoso} model caused by the lack of articulations. This is reasonable because simpler pieces generally have fewer articulations. However, it may induce a biased piece classification in some instances.
    

\begin{figure}[h!]
\centering
\subfloat[a]{
  \includegraphics[clip,width=0.6\columnwidth]{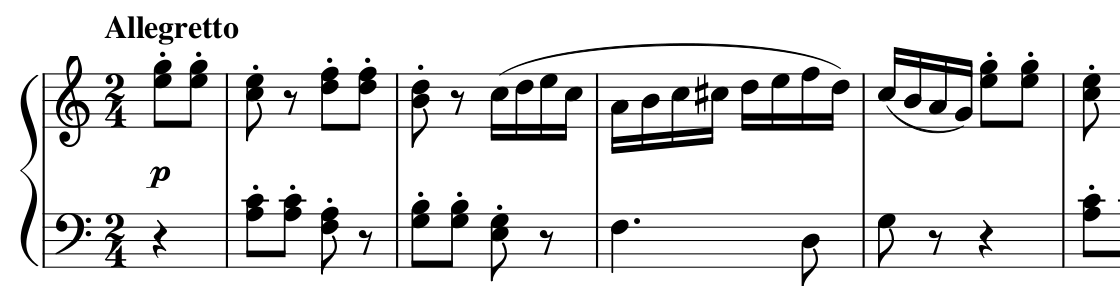}
}

\subfloat[b]{
  \includegraphics[clip,width=0.6\columnwidth]{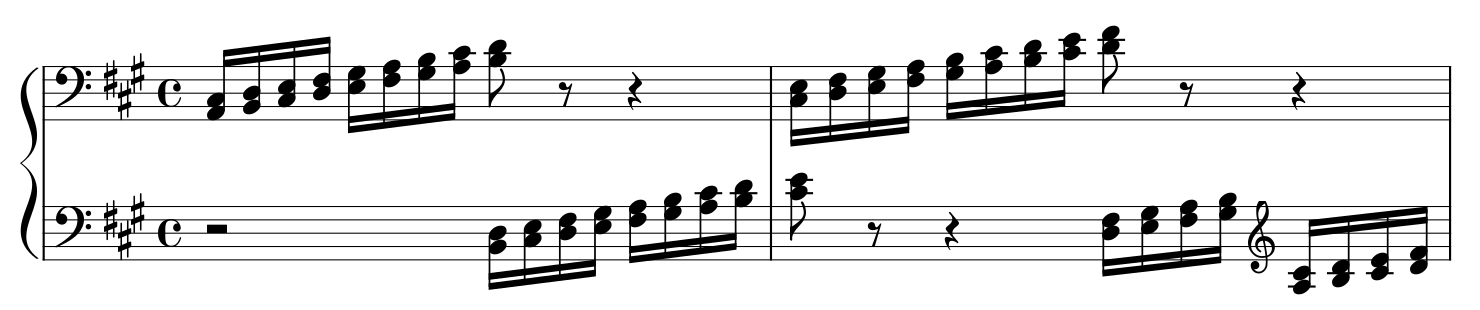}
}

\caption{Example of under-performance of the model trained on \emph{x} (a) Fragment of the piece \emph{3rd movement (Rondo) from Piano Sonata (Facile), C major KV 545, W. A. Mozart}. (b) Fragment of the piece \emph{Exercise 2a WoO 6,2a, Johannes Brahms}.}
\label{fig:case4}
\end{figure}

Lastly, in Figure~\ref{fig:case4}, we highlight the limitations of \emph{x} representation. The first example is \emph{3rd movement (Rondo) from Piano Sonata (Facile), C major KV 545, W. A. Mozart}. The ground truth label is 4, and the ensemble model's prediction is 5. The prediction made by the \emph{argnn} model is 4, while the \emph{pitch} model predicts 3, and the \emph{virtuoso} model predicts 7. The second example is \emph{Exercise 2a WoO 6,2a, Johannes Brahms}. The ground truth label is 5, and the ensemble model's prediction is 5. The prediction made by the \emph{argnn} model is 8, while the \emph{pitch} model predicts 3 and the \emph{virtuoso} model predicts 5. In both cases, we can observe that the model trained on \emph{pitch} predicts lower grades than the other models. We think it is because \emph{pitch} representation has limitations in understanding polyphony, as we state on \cite{ramoneda2022}. Finally, in the second example, we want to emphasize the strong prediction of the \emph{argnn} model. Considering the technical nature of \emph{Exercise 2a WoO 6,2a, Johannes Brahms}, this is reasonable.


The case study showed that the three main input representations contribute to the final \emph{ensemble} model. However, the annotations of Henle Verlag only provide information about the general difficulty of a piece, not the technical, expressiveness, or structural difficulty, making the models trained in a particular dimension have an unreliable prediction. The study exposed some limitations of the methods proposed and demonstrated the relationship between piano performance and input representations.

%% file: 7_conclusion.tex
\section{Challenges}
\label{sec:challenges}

Our results confirm a correlation between the difficulty of piano performance and various dimensions of musical performance, guided by prior musicology research~\cite{cook1999analysing}. This leads to several avenues for further research.

\textbf{Training with fragments of chunks.}
The scarcity of labeled data at the piece level for score difficulty analysis has led us to explore training with smaller fragments instead of the entire piece in Section~\ref{exp:fragments}. Although there are many weak annotations regarding the number of labeled segments, using fragments to train the model is not straightforward. Further research in musicology and computational musicology is necessary to understand how fragments contribute to the overall difficulty of a piece. This understanding will be crucial for future advancements in difficulty performance analysis.

\textbf{Training simultaneously the backbone models and the difficulty levels.} Human pianists progressively in difficulty learn the challenging physical aspects of pieces and their expressive rendition, i.e., training together automatic piano fingering, expressive piano performance, and score difficulty classification. This suggests that models might benefit from parallel training, as seen in multi-task learning, or from a sequential and cyclic approach, continual learning, to enhance their capabilities. Nonetheless, a key challenge is the absence of a music score dataset with label annotations covering difficulty, fingering, and expressivity simultaneously. Enhancements in models for the three tasks mentioned are crucial for generating synthetic data to bridge this alignment gap.

\textbf{Better performance generation models. } We have modeled performance through automatic piano fingering~\cite{ramoneda2022automatic} and expressive piano performance generation~\cite{jeong2019virtuosonet}. However, both of these tasks are yet to be fully resolved, and improvements in their performance may enhance the analysis of performance from the score, particularly for the task of piano difficulty classification. In addition, as is shown in the case study, Section~\ref{exp:examples}, increasing the robustness of performance generation models in the absence of slurs or dynamics might significantly enhance the accuracy in evaluating the difficulty levels of musical pieces.

\textbf{Data augmentation.} The application of data augmentation is fundamental in tasks with very little data, such as the ones presented in the present research. However, data augmentation techniques traditionally used on symbolic music tasks~\cite{nestor2021augmented,yang2022large} can not be directly applied to the difficulty classification task. For instance, transposition augmentation could alter the distances between the black and white notes of the piano, creating a very different technical difficulty. Also, a random combination of the fragments of the pieces may cause significant changes of difficulty in the union between the fragments. Furthermore, the exclusion of parts of the pieces can lead to the omission of valuable information about the difficulty of the musical work.

\textbf{Learning with noisy labels.} In future work, it is crucial to utilize crowdsourced annotations about difficulty from websites such as 8notes to improve performance on the \emph{CIPI} dataset. Additionally, exploring how to expand the \emph{CIPI} dataset domain through self-supervision in large corpora and semi-supervision is a promising research avenue.

\textbf{Using all the multi-modal data available.} The information of classical performances can be represented in multiples modalities: symbolic piano-roll, symbolic scores (used in \emph{Mirkoskosmos-difficulty} and \emph{CIPI}, pdf scores, audio, and video of the performance. Although the symbolic score modality is a well-starting point because of interpretability, the other modalities have other advantages, such as implicitly containing the technique and expressive information. Exploring how to analyze the performance difficulty in other modalities may be helpful.

\textbf{Multi-ranking.} The classification of the most difficult musical pieces can be found from various sources such as other music systems, publishers, or examination boards in addition to Henle Verlag. The most time-consuming task of the present research was compiling a high-quality collection of symbolic music. However, searching other sources can clarify the ranking of the collected \emph{CIPI} musical works. We believe that having multiple perspectives on the concept of difficulty can aid in finding a more objective view of musical performance difficulty.

\vspace{-0.5cm}
\section{Conclusions}
\label{sec:conclusion}

In this work, we introduced a new dataset of symbolic piano scores with difficulty level annotations from the recognized classical music publisher Henle Verlag with a new methodology to create MIR datasets. We curated the \emph{CIPI} dataset after evaluating and rectifying the automatic matching between public domain scores and Henle Verlag annotations by an expert pianist. 
We trained models based on various dimensions of musical performance on \emph{CIPI} dataset, inspired by prior musicology research and in comparison with the \emph{Mikrokoskmos-difficulty} dataset. Following the approach outlined in~\cite{cook1999analysing}, we combined the predictions of multiple models trained on different musical performance dimensions, which resulted in improved performance compared to individual models. Our models achieved a balanced accuracy of 39.5\% and a median square error of 1.1 across the nine difficulty levels in the CIPI dataset. We emphasized the importance of choosing the appropriate loss function for the ordinal classification task in training on the CIPI dataset. Additionally, we conducted extensive experiments to inform further research, including training with fragments of pieces instead of whole pieces, using different methods for feature fusion, limiting the classes in the \emph{CIPI} dataset to only 3, and training only with the shortest pieces. 
We conclude that difficulty analysis is a very challenging and complex task involving many dimensions. 

With this paper, we want to lay the foundations for research on difficulty analysis in piano repertoire from a MIR perspective. Advancing the research for better structuring extensive collections of classical music to increase the diversity of the piano curriculum on music education and enhancing the participation of the student on the election of the mentioned curriculum.
Furthermore, studying the difficulty analysis through computational approaches contributes to the research on automatic music arrangement systems and other automatic composition tools for music education. In addition, the task we want to establish with this paper may help design curriculum learning strategies for other tasks, such as automatic piano fingering, automatic music generation, or expressive performance.
In future work, we plan to create more explainable representations for computational musicology-oriented research and explore other data sources to classify difficulty.